\documentclass[aps,pra,twocolumn]{revtex4-1}

\usepackage[english]{babel}
\usepackage[T1]{fontenc}
\usepackage[utf8]{inputenc}

\usepackage{bbm}  
\usepackage{graphicx}
\usepackage{amsmath}
\usepackage{amssymb}
\usepackage{braket}
\usepackage{booktabs}
\usepackage{simplewick}
\usepackage{bm} 

\usepackage{cellspace}
\newcommand\cincludegraphics[2][]{\raisebox{-0.4\height}{\includegraphics[#1]{#2}}}

\usepackage[autostyle=true]{csquotes}

\begin{document}                                                                                                                                                                   
                                                                                                                                                                                   
\title{Automatic Routing of Goldstone Diagrams using Genetic Algorithms}
\author{Nils Herrmann}
\email{N.Herrmann@uni-koeln.de}
\author{Michael Hanrath}
\email{Michael.Hanrath@uni-koeln.de}

\affiliation{%
Institute for Theoretical Chemistry, University of Cologne, 
Greinstra\ss e 4, 50939 Cologne, Germany
}%

\date{\today}

\begin{abstract}
This paper presents an algorithm for an automatic transformation (=routing) of time ordered
topologies of Goldstone diagrams 
(i.e. Wick contractions) into graphical representations of these topologies.
Since there is no hard criterion for an optimal routing, the proposed algorithm minimizes
an empirically chosen cost function over a set of parameters. Some of the latter are naturally
of discrete type (e.g. interchange of particle/hole lines due to antisymmetry) while others
(e.g. $x,y$-position of nodes) are naturally continuous. In order to arrive at a manageable 
optimization problem the position space is artificially discretized. In terms of the 
(i) cost function, (ii) the discrete vertex placement, (iii) the interchange of particle/hole lines
the routing problem is now well defined and fully discrete.
However, it shows an exponential complexity with the number of vertices
suggesting to apply a genetic algorithm for its solution. The presented algorithm is capable
of routing non trivial (several loops and crossings) Goldstone diagrams. The resulting diagrams
are qualitatively fully equivalent to manually routed ones.
The proposed algorithm is successfully applied to several Coupled Cluster approaches and
a perturbative (fixpoint iterative) CCSD expansion with repeated diagram substitution. \\
\end{abstract}

\maketitle


\section{Introduction}
\label{introduction}

Coupled-Cluster (CC) theory was originally formulated in the field of 
nuclear physics by Coester and K\"ummel \cite{Coester_1958, Coester_2_1958}. The development of diagrammatic 
representations of the underlying terms was pioneered by \v{C}\'{i}\v{z}ek and Paldus \cite{Cizek_1966, Paldus_1975, Paldus_1977}. 
These representations were based on the original diagrammatic formulations of Goldstone \cite{Goldstone_1957} or 
Hugenholtz \cite{Hugenholtz_1957}, who generalized Feynman's diagrammatic representation of particle 
interactions \cite{Feynman_1949}, for the 
many-body perturbation series. 

The efficient generation of Hugenholtz- or Goldstone-type diagrams is an 
ongoing task of computational quantum chemistry and physics, where plenty of algorithms have been proposed 
in the literature (see e.g. \cite{Paldus_1973, Csepes_1988, Lyons_1994, Derevianko_2002, Mathar_2007}). 
Due to tremendous difficulties in the derivation of higher order ($5,6,\ldots$) CC equations, using 
algebraic terms and Wick's theorems \cite{Wick_1950}, fast equation derivation is one of the key applications 
of diagrammatic CC techniques \cite{Kallay_2001}, where a lot of progress has been made in the last 
20 years (see e.g. \cite{Harris_1999, Kallay_2001, Dzuba_2009}). 

With the advancing algorithm progress deriving CC equations from diagrammatic representations, more and more 
complex structures in storing and processing the latter representations were created, while the original 
intuitively understandable diagrammatic structures were lost. These structures however, could be of 
great importance in the fundamental CC research by supplying insight into more complex contraction 
patterns or by enabling a sophisticated analyzing tool for new CC approaches.

Considering CC diagrams of higher orders, a large variety of possibilities to correctly illustrate (route) all diagrammatic fragments exists. 
In particular diagrams, which are not present in the original quantum chemistry CC theory (e.g. including 
powers of the Hamiltonian ($\hat{H}_N^2, \ldots$) or three particle interrelating operators ($\hat{W}_N$)), possess 
neither unique nor well defined optimal layouts. 
Finding these can be a very complicated task when done by hand.

In this paper, an automatic Goldstone diagram optimization algorithm (\glqq{A}utoDiag\grqq{}) 
is presented. In the current version, it is capable of transforming arbitrary Goldstone diagram 
topologies to an optimized, readable diagram layout (its representation) meeting a set of (hard) physical and (soft) optical constraints.

\begin{figure}
\includegraphics[scale=1]{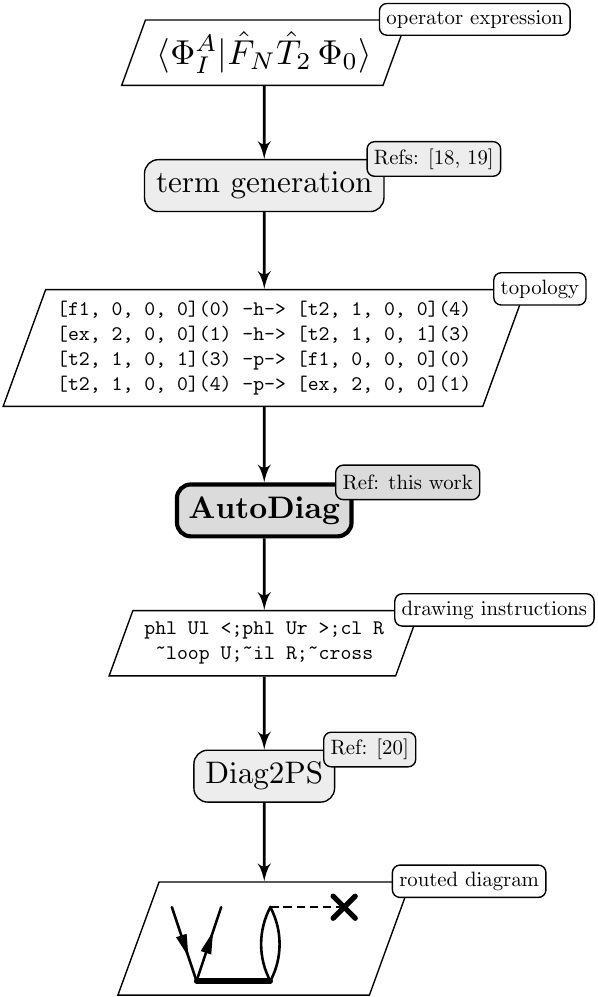}
\caption{\label{AutoDiag.fig}Execution flow chart diagram of the term generation engine \enquote{sqdiag} \cite{sqdiag1, sqdiag2}, the automatic routing tool \enquote{AutoDiag} and the text driven 
diagram assembling tool \enquote{Diag2PS} \cite{Diag2PS} explicitly showing input and output data.}
\end{figure}

As explicitly illustrated in figure \ref{AutoDiag.fig}, diagram topologies are generated from operator expressions using the term 
generation engine \enquote{sqdiag} \cite{sqdiag1, sqdiag2} while diagrammatic layouts are plotted using the text driven diagram assembling 
tool \enquote{Diag2PS}. \cite{Diag2PS} 

\section{Theory}
\label{theory}

In this section, the theoretical basis for CC and its representation using diagrammatic techniques is briefly
introduced (for further information see e.g. \cite{Shavitt_Book, Harris_Book, Crawford_Book}).

\subsection{Many body operators in second quantized form}

In the CC ansatz, the exact many-particle 
wavefunction $\ket{\Psi}$ is obtained via the application of the wave operator $e^{\hat{T}}$ to the Hartree-Fock (HF)
reference determinant $\ket{\Psi_0}$.
Here, $\hat{T}$ denotes the total cluster operator with 
\begin{equation}
\label{totalT.eq}
\hat{T} = \hat{T}_1 + \hat{T}_2 + \ldots,
\end{equation}

where the individual T$_\nu$ operators denote $\nu$-particle substitutions via 
\begin{equation}
\label{singleT.eq}
\hat{T}_\nu = \frac{1}{(\nu!)^2}\sum_{\substack{i_1\ldots i_\nu \\ a_1\ldots a_\nu}} t_{i_1\ldots i_\nu}^{a_1\ldots a_\nu}\hat{a}_{a_1}^\dagger\ldots \hat{a}_{a_{\nu}}^\dagger \hat{a}_{i_{\nu}}\ldots\hat{a}_{i_{1}} \;.
\end{equation}

Using the common index notations for occupied ($\mathbb{O}$) and virtual ($\mathbb{V}$) orbital 
spaces
\begin{align*}
i,j,k,\ldots &\in \mathbb{O} \\
a,b,c,\ldots &\in \mathbb{V} \\
p,q,r,\ldots &\in \mathbb{O}\cup\mathbb{V}\;\text{,}
\end{align*}

the normal ordered Hamiltonian $\hat{H}_N$ takes the form 
\begin{align*}
\hat{H}_N &= \hat{F}_N + \hat{V}_N \\
&= \sum_{pq} f^p_q \hat{a}_{p}^{\dagger}\hat{a}_{q} + \frac{1}{4}\sum_{pqrs}v^{pq}_{rs} \hat{a}_{p}^{\dagger}\hat{a}_{q}^{\dagger}\hat{a}_{s}\hat{a}_{r}\;,
\end{align*}

where $f_q^p$ and $v^{pq}_{rs}$ denote the one- and two-particle integrals
\begin{align*}
f_q^p &= \braket{p | \hat{f}(1) | q} \text{ and }\\
v_{rs}^{pq} &= \braket{pq|\hat{g}(1,2)|rs} - \braket{pq|\hat{g}(1,2)|sr}\;,
\end{align*}
respectively. In general, the amplitudes $t$ and the integrals $v$ are antisymmetric with respect to index permutations such that
\begin{alignat}{3}
\label{antisym1.eq}
t_{pq\ldots}^{rs\ldots} &= -t_{pq\ldots}^{sr\ldots} &&= -t_{qp\ldots}^{rs\ldots} &&= t_{qp\ldots}^{sr\ldots} = \ldots \\
v_{pq}^{rs} &= -v_{pq}^{sr} &&= -v_{qp}^{rs} &&= v_{qp}^{sr}\;.
\label{antisym2.eq}
\end{alignat}

\subsection{Algebraic evaluation}
\label{Algebraic}

The projected and similarity transformed CC equations are given by
\begin{align}
\label{CCEnergy.eq}
\braket{\Psi_{0}| e^{-\hat{T}}\hat{H}_Ne^{\hat{T}} \Psi_0} &= E_{\text{corr}} \\
\braket{\Psi_{I\ldots}^{A\ldots}| e^{-\hat{T}}\hat{H}_Ne^{\hat{T}} \Psi_0} &= 0\;,
\label{CCAmplitude.eq}
\end{align}
where capital indices $A\ldots$ and $I\ldots$ denote external projections. 

The similarity transformed Hamiltonian $\bar{H} = e^{-\hat{T}}\hat{H}_Ne^{\hat{T}}$ can be expanded in the Baker-Campbell-Hausdorff (BCH) 
formula \cite{Baker_1902, Campbell_1897, Hausdorff_1906} to yield nested commutators of $\hat{H}_N$ and $\hat{T}$, which naturally vanish for  
quintuply (and higher) nested commutators due to the two-particle interrelating nature of $\hat{H}_N$. 
Inserting the BCH series of $\bar{H}$ into 
the CC equations (\ref{CCEnergy.eq}) and (\ref{CCAmplitude.eq}), only fully contracted Wick-terms survive.
Furthermore, all participating operators from $\hat H_N$ must be at least singly contracted to all $\hat{T}$ operators on its right.

As an example for algebraic evaluation consider the term 
\begin{equation}
\label{example.eq}
\contraction[1ex]{E_\text{corr} \quad \longleftarrow \quad \langle\Psi_0 | }{\hat{V}}{_N}{\hat{T}}
\contraction[1.5ex]{E_\text{corr} \quad \longleftarrow \quad \langle\Psi_0 | }{\hat{V}}{_N}{\hat{T}}
\contraction[2ex]{E_\text{corr} \quad \longleftarrow \quad \langle\Psi_0 | }{\hat{V}}{_N}{\hat{T}}
\contraction[2.5ex]{E_\text{corr} \quad \longleftarrow \quad \langle\Psi_0 | }{\hat{V}}{_N}{\hat{T}}
E_\text{corr} \quad \longleftarrow \quad \langle\Psi_0 | \hat{V}_N\hat{T}_2\Psi_0\rangle\;,
\end{equation}

which contributes to the correlation energy (\ref{CCEnergy.eq}). To algebraically evaluate (\ref{example.eq}), all possible full contraction 
patterns of the underlying second quantized operator strings must be collected. These result in products of Kronecker-deltas, which bring down the 
general index summation ($p,q,r,s$) by binding them to particle/hole indices ($i,j,a,b$) of the amplitudes $t$. Using antisymmetry, all redundant 
two-particle integrals $v$ can then be merged to end up with
\begin{multline*}
\contraction[1ex]{\langle\Psi_0 | }{\hat{V}}{_N}{\hat{T}}
\contraction[1.5ex]{\langle\Psi_0 | }{\hat{V}}{_N}{\hat{T}}
\contraction[2ex]{\langle\Psi_0 | }{\hat{V}}{_N}{\hat{T}}
\contraction[2.5ex]{\langle\Psi_0 | }{\hat{V}}{_N}{\hat{T}}
\langle\Psi_0 | \hat{V}_N\hat{T}_2\Psi_0\rangle 
= \frac{1}{16}\sum_{pqrs}\sum_{ijab} v^{pq}_{rs}t_{ij}^{ab} \;\cdot \\
\left[\rule{0pt}{20pt}\right.
\contraction[1.0ex]{\{\hat{a}_p^\dagger \hat{a}_q^\dagger \hat{a}_s }{\hat{a}}{_r\} \{}{\hat{a}}
\contraction[1.5ex]{\{\hat{a}_p^\dagger \hat{a}_q^\dagger }{\hat{a}}{_s \hat{a}_r\} \{\hat{a}_a^\dagger }{\hat{a}}
\contraction[2.0ex]{\{\hat{a}_p^\dagger }{\hat{a}}{_q^\dagger \hat{a}_s \hat{a}_r\} \{\hat{a}_a^\dagger \hat{a}_b^\dagger }{\hat{a}}
\contraction[2.5ex]{\{}{\hat{a}}{_p^\dagger \hat{a}_q^\dagger \hat{a}_s \hat{a}_r\} \{\hat{a}_a^\dagger \hat{a}_b^\dagger \hat{a}_j }{\hat{a}}
\{\hat{a}_p^\dagger \hat{a}_q^\dagger \hat{a}_s \hat{a}_r\} \{\hat{a}_a^\dagger \hat{a}_b^\dagger \hat{a}_j \hat{a}_i\} \quad + 
\contraction[1.0ex]{\{\hat{a}_p^\dagger \hat{a}_q^\dagger \hat{a}_s }{\hat{a}}{_r\} \{\hat{a}_a^\dagger }{\hat{a}}
\contraction[1.5ex]{\{\hat{a}_p^\dagger \hat{a}_q^\dagger }{\hat{a}}{_s \hat{a}_r\} \{}{\hat{a}}
\contraction[2.0ex]{\{\hat{a}_p^\dagger }{\hat{a}}{_q^\dagger \hat{a}_s \hat{a}_r\} \{\hat{a}_a^\dagger \hat{a}_b^\dagger }{\hat{a}}
\contraction[2.5ex]{\{}{\hat{a}}{_p^\dagger \hat{a}_q^\dagger \hat{a}_s \hat{a}_r\} \{\hat{a}_a^\dagger \hat{a}_b^\dagger \hat{a}_j }{\hat{a}}
\{\hat{a}_p^\dagger \hat{a}_q^\dagger \hat{a}_s \hat{a}_r\} \{\hat{a}_a^\dagger \hat{a}_b^\dagger \hat{a}_j \hat{a}_i\} \quad + \\
\contraction[1.0ex]{\{\hat{a}_p^\dagger \hat{a}_q^\dagger \hat{a}_s }{\hat{a}}{_r\} \{}{\hat{a}}
\contraction[1.5ex]{\{\hat{a}_p^\dagger \hat{a}_q^\dagger }{\hat{a}}{_s \hat{a}_r\} \{\hat{a}_a^\dagger }{\hat{a}}
\contraction[2.0ex]{\{\hat{a}_p^\dagger }{\hat{a}}{_q^\dagger \hat{a}_s \hat{a}_r\} \{\hat{a}_a^\dagger \hat{a}_b^\dagger \hat{a}_j }{\hat{a}}
\contraction[2.5ex]{\{}{\hat{a}}{_p^\dagger \hat{a}_q^\dagger \hat{a}_s \hat{a}_r\} \{\hat{a}_a^\dagger \hat{a}_b^\dagger }{\hat{a}}
\{\hat{a}_p^\dagger \hat{a}_q^\dagger \hat{a}_s \hat{a}_r\} \{\hat{a}_a^\dagger \hat{a}_b^\dagger \hat{a}_j \hat{a}_i\} \quad + 
\contraction[1.0ex]{\{\hat{a}_p^\dagger \hat{a}_q^\dagger \hat{a}_s }{\hat{a}}{_r\} \{\hat{a}_a^\dagger }{\hat{a}}
\contraction[1.5ex]{\{\hat{a}_p^\dagger \hat{a}_q^\dagger }{\hat{a}}{_s \hat{a}_r\} \{}{\hat{a}}
\contraction[2.0ex]{\{\hat{a}_p^\dagger }{\hat{a}}{_q^\dagger \hat{a}_s \hat{a}_r\} \{\hat{a}_a^\dagger \hat{a}_b^\dagger \hat{a}_j }{\hat{a}}
\contraction[2.5ex]{\{}{\hat{a}}{_p^\dagger \hat{a}_q^\dagger \hat{a}_s \hat{a}_r\} \{\hat{a}_a^\dagger \hat{a}_b^\dagger }{\hat{a}}
\{\hat{a}_p^\dagger \hat{a}_q^\dagger \hat{a}_s \hat{a}_r\} \{\hat{a}_a^\dagger \hat{a}_b^\dagger \hat{a}_j \hat{a}_i\} 
\left.\rule{0pt}{20pt}\right] \\
= \frac{1}{16}\sum_{ijab} \left[ v^{ij}_{ab} - v^{ij}_{ba} - v^{ji}_{ab} + v^{ji}_{ba}\right]t_{ij}^{ab} 
= \frac{1}{4}\sum_{ijab}v^{ij}_{ab}t_{ij}^{ab}\;.
\end{multline*}

As is clearly recognized, even this fairly simple algebraic derivation example turns out to be annoyingly complex. In part, this complexity arises from 
the fact that the Wick theorems do not incorporate the antisymmetry of the tensors $v$ and thereby produce a lot of redundant terms. 

\subsection{Diagrammatic evaluation}

A different approach, pioneered by \v{C}\'{i}\v{z}ek and Paldus \cite{Cizek_1966, Paldus_1975, Paldus_1977}, was found when using diagrammatic 
techniques to derive the CC equations. In the particle/hole formalism (derived and applied also in the algebraic framework), the definition of all second quantized operators acting on $\mathbb{O}$
is inverted to yield the \glqq{q}uasi-particle operators\grqq{} $b$:
\begin{align*}
\forall_{i \in \mathbb{O}}&
\begin{cases}
\hat{a}_i^\dagger = \hat{b}_i & \text{ hole annihilator}\\
\hat{a}_i = \hat{b}_i^\dagger & \text{ hole creator}
\end{cases} \\
\forall_{a \in \mathbb{V}}&
\begin{cases}
\hat{a}_a^\dagger = \hat{b}_a^\dagger & \text{ particle creator} \\
\hat{a}_a = \hat{b}_a & \text{ particle annihilator}
\end{cases}
\end{align*}

\begin{figure}
\includegraphics[scale=0.7]{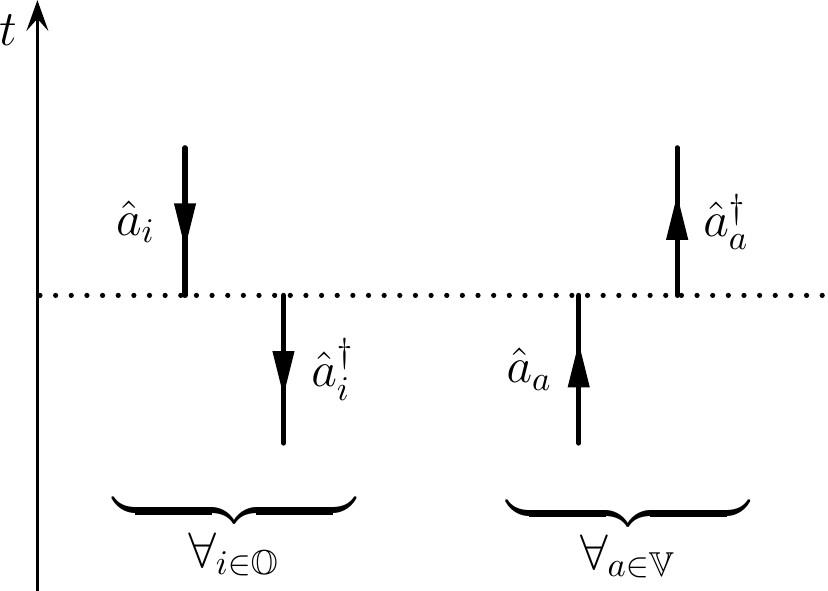}
\caption{\label{phlines.fig} Illustration of second quantized operators as particle/hole lines in a Feynman-type space-time diagram.}
\end{figure}

These may now be illustrated by directed lines in a Feynman-type space-time diagram (c.f. fig. \ref{phlines.fig}).

Here, hole-operators (acting on $\mathbb{O}$) are represented as directed lines facing down (moving backwards in time) and particle-operators 
(acting on $\mathbb{V}$) are represented as directed lines facing up (moving with time). A further distinction is made for creators 
and annihilators with respect to a certain event in time (e.g. a particle interrelating operator contracted to the individual particle/hole 
lines). All outgoing particle/hole lines represent creators ($\hat{a}_i^\dagger$ and $\hat{a}_a^\dagger$) and all incoming particle/hole lines represent
annihilators ($\hat{a}_i$ and $\hat{a}_a$). 

The most important feature of the particle and hole line representation is found when considering Wick-contractions. The 
only surviving contraction types are 
\begin{align}
\contraction{}{\hat{a}}{_i^\dagger }{\hat{a}}
\hat{a}_i^\dagger \hat{a}_j &= \delta_{ij} \\
\contraction{}{\hat{a}}{_a }{\hat{a}}
\hat{a}_a \hat{a}_b^\dagger &= \delta_{ab}\;.
\end{align}

\begin{figure}
\tabcolsep=40pt
\begin{tabular}{cc}
\includegraphics[scale=0.6]{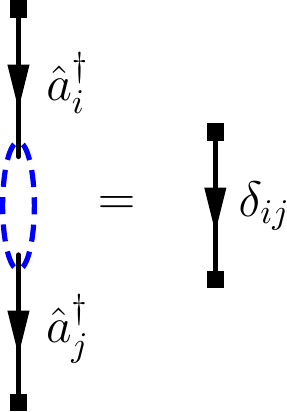}
&
\includegraphics[scale=0.6]{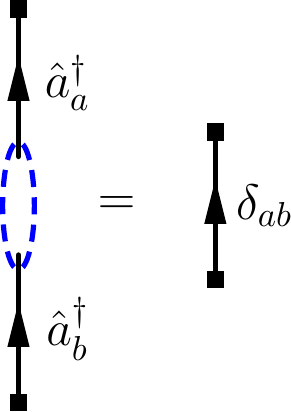}
\\
(a) &(b)
\end{tabular}
\caption{\label{contraction.fig}
Illustration of the incorporated contraction rules for hole (a) and particle (b)
lines.}
\end{figure}

As it turns out, these contraction rules are perfectly satisfied when connecting particle/hole lines heads to tails (c.f. fig. \ref{contraction.fig}). 
First of all, it is only possible to connect two arrows if both of them are facing up or facing down allowing for contractions among two holes or two 
particles only. Secondly, an outgoing line must always be connected with an incoming line thus only ensuring contractions among creators and annihilators.

\begin{figure}
\includegraphics[scale=0.98]{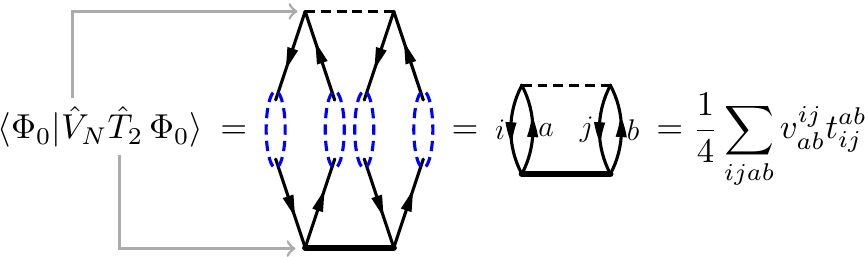}
\caption{\label{Connect.fig} Schematic evaluation of an examplatory CC integral contributing to the correlation energy using Goldstone diagrammatic techniques.}
\end{figure}

To fully represent particular many-body integrals of the CC equations (\ref{CCEnergy.eq}) and (\ref{CCAmplitude.eq}) -- as e.g. illustrated in figure \ref{Connect.fig} -- the occuring particle interrelating 
operators of the Hamiltonian $\hat{H}_N$ and the cluster operator $\hat{T}$ need to be represented. In general, particle interrelating operators are illustrated 
by horizontally arranged vertices connected by a dashed interaction line, which are connecting several particle/hole lines.

The individual cluster operators $\hat{T}_\nu$ have fixed particle/hole indices and always promote particles from $\mathbb{O}$ to $\mathbb{V}$. Therefore, they 
always possess the same \glqq{V}\grqq{-}shape of connected particle/hole lines. Cluster operators $\hat{T}_\nu$ with $\nu \ge 2$ are illustrated by $\nu$
horizontally arranged vertices connected by a bold cluster line. 

In figure \ref{Connect.fig}, the same example (eq. \ref{example.eq}) as evaluated in section \ref{Algebraic} is illustrated using the corresponding 
Goldstone representations. From the definition of the integral, the participating operator representations (with their correct ranks) may be 
extracted and illustrated  
according to their time of operator application (with respect to the ket Fermi-vaccuum $\ket{\Psi_0}$) in a vertically 
advancing time scale. Connecting all particle/hole lines as mentioned above (heads to tails) leads to the complete Goldstone diagram, which consists 
of two particle/hole line loops in this example. This diagram is completely translatable (including sign and prefactor) to the algebraic evaluation of the integral.


\section{Discrete optimization problem: Diagram routing} 
\label{OptimizationProblem}

Since there is no hard criterion for an optimal Goldstone diagram layout, a manageable global optimization problem of type
\begin{equation}
\min_{\vec{g}(\vec{x}) = \vec{0}}f(\vec{x}) \text{  with  } f:\mathbb{G}^n \to \mathbb{R}
\end{equation}
is defined. Here, $\vec{x}$ represents the target quantity, wich completely defines one specific layout (e.g. all information of solution space $\mathbb{G}$ to fully 
characterize all $n$ vertices).

In order to guarantee physically correct Goldstone diagrams, i.e. a conserved operator time ordering, specific constraints $\vec{g}(\vec{x})$ are placed 
upon the minimalization problem (c.f. subsection \ref{Constraints}). This leads to a significantly reduced area of $\bar{\mathbb{G}}^n \subseteq \mathbb{G}^n$ where all constraints are fulfilled.
This remaining area represents the solution space the algorithm should traverse.

Finally, for each layout $\vec{x}$ a certain cost value $f(\vec{x})$ (c.f. subsection \ref{CostEvaluation}) is defined, such that the global minimum of $f$ 
represents an optimally routed diagram. 

\begin{figure}
\begin{minipage}{0.43\columnwidth}
\includegraphics[scale=0.74]{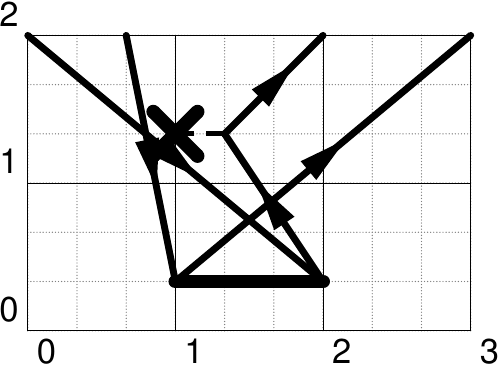}
\end{minipage}
\begin{minipage}{0.1\columnwidth}
$\longrightarrow$
\end{minipage}
\begin{minipage}{0.43\columnwidth}
\includegraphics[scale=0.74]{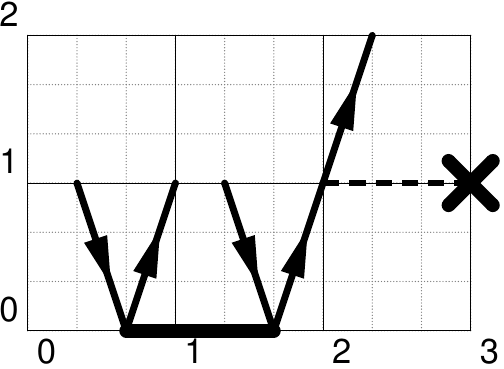}
\end{minipage}
\caption{\label{grid.fig}An example to the high-dimensional and discrete Goldstone diagram optimization problem.}
\end{figure}

\subsection{Explicit spatial discretization}
\label{Discretization}

The automatic routing of Goldstone diagrams clearly resembles a high-dimensional problem, since all vertices of the latter must be placed in their optimal position. 
Therefore, two dimensional diagrams without any 
constraints possess a dimensionality of $2n$ (for $n$ vertices with $x$ and $y$ coordinates). Per definition however, the optimization problem is not discrete. 
Only when introducing a discrete solution space for each 
diagram, an optimization algorithm can perform well. 
This work employs the text driven diagram assembling tool \glqq{D}iag2PS\grqq{} \cite{Diag2PS}, which is capable of 
illustrating Goldstone, Hugenholtz or classical Feynman diagrams. 
In this tool, a discrete grid of thirds is used to place the vertices of the assembled diagrams. This grid was used to discretize the vertex positions in this work. 

An example illustration of the grid-based Goldstone diagram optimization problem is shown in figure \ref{grid.fig}.
The diagram on the left hand side was generated randomly following its time-ordered topology from the CCSD expansion,  
to illustrate a random starting point of the optimization. On the right hand side, the desired minimum is shown. Each vertex position is exactly 
characterized by its integer $x$ and $y$ coordinates on the grid.

\subsection{PHL structure}
\label{PHLStructure}

Knowledge about the coordinates $x$ and $y$ alone is not sufficient 
to fully characterize Goldstone diagrams. It is also mandatory to optimize the topology, i.e. the particle/hole line (PHL) structure, of the diagram. 
This structure is composed in the following fashion:
\begin{itemize}
\item[(i)] A single PHL is defined as a directed line connecting two distinct vertices. The spatial extent (curvature, straightness) of the latter is arbitrary as long as the 
direction (facing up or facing down) as well as the starting and ending vertices of the PHL are conserved.
\item[(ii)] Several PHLs may form a connected segment (a path) of alternating incoming and outgoing PHLs advancing over several vertices (c.f. fig \ref{PHLSegments.fig}). 
Therefore, every PHL segment can either start and end at exactly the same vertex (representing 
a loop, c.f. segments (a) and (b)) or start and end at different external vertices (c.f. segment (c)), which stop the connected path. 
\begin{figure}
\includegraphics[width=\columnwidth]{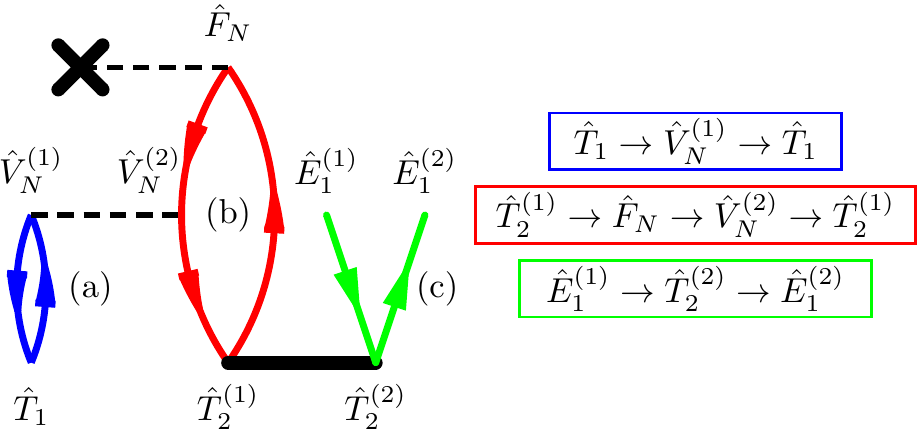}
\caption{\label{PHLSegments.fig} Illustration of different PHL segments from an examplatory Goldstone diagram.}
\end{figure}
\item[(iii)] The collection of all PHL segments of a Goldstone diagram is denoted by its PHL structure, which fully characterizes its topology. 
\end{itemize}

The PHL structure of any Goldstone diagram is naturally discrete
since (i) all PHLs must start and end at specific single vertices and (ii) all incoming or outgoing lines are uniquely mapped to exactly one vertex (i.e. 
no vertex can be connected by more than one incoming or outgoing line).

\subsection{Constraints}
\label{Constraints}

In case of Goldstone diagrams there are several constraints for a physically correct representation of the latter. 
In summary, the following constraints are placed on the optimization problem:
\begin{itemize}
\item[(i)] Every interaction or cluster vertex/line (representing one specific particle interrelating operator from the Hamiltonian or one specific cluster operator) is 
identified with one distinct relative position in time. This time scale is based on the order of operator applications to the ket Fermi vaccuum $\ket{\Psi_0}$, 
where first acting operators are illustrated earlier in time then last acting operators. In this work, the time axis in Goldstone diagrams is illustrated 
as advancing from the bottom to the top.
\item[(ii)] Since interaction or cluster lines can consist of several vertices (and each interaction or cluster line is only identified with one time position), all 
those vertices must possess the same time position and must therefore be illustrated horizontally arranged. 
\item[(iii)] Whenever two operators commute (e.g. several $\hat{T}$ operators for disjoint $\mathbb{O}$ and $\mathbb{V}$) they are capable to
interchange their time position arbitrarily. Conventionally, such operators are illustrated at the same time position to illustrate their interchangeability. 
\item[(iv)] An exception is made for all external lines originating from projections. In principle, external projections always possess the highest position in virtual time since they are 
always acting from the left ($\bra{\Psi_{I\ldots}^{A\ldots}}\hat{X} = \bra{\Psi_0}\hat{a}_I^\dagger\ldots \hat{a}_A\hat{X}$). However, since they are represented by isolated particle/hole 
lines, they are allowed to fall down to the individual cluster or interaction lines they are connected to. As a hard constraint they must never fall below their connected interaction 
or cluster lines.
\end{itemize}

Taking all of these restrictions for granted, the resulting Goldstone diagrams can always be correctly translated to their algebraic form. With increasing diagram size 
however, the number of physically correct representations increases enormously while only few are still intuitively readable (analyzable). This does not only include the 
spatial positioning of all vertices but also the PHL structure. 

\subsection{Degrees of freedom}
\label{DegreesOfFreedom}

If all constraints are fulfilled, the following degrees of freedom allow for an optimization of the cost:

\begin{itemize}
\item[(i)] The horizontal axis of a Goldstone diagram has no algebraic implications. If a vertex is illustrated more to the left or to the right for instance, has no effect on the 
algebraic correctness of the diagram. 
\item[(ii)] The actual position in time of any Goldstone interaction or cluster line is arbitrary as long as it does not violate the time ordering itself. This means that vertices 
may be placed closer together or further apart in the vertical time dimension without interchanging their position. 
\item[(iii)] Due to the antisymmetry of the included tensors (c.f. equations \ref{antisym1.eq} and \ref{antisym2.eq}),
pairs of two outgoing (creators) or two incoming (annihilators) PHLs connected to the same interaction or cluster line may be interchanged. This changes the topology of the 
Goldstone diagram and may therefore lead to a significant 
change in the PHL structure, which e.g. includes the number of loops of a diagram (c.f. fig. \ref{Antisym.fig}). 
\begin{figure}
\includegraphics[scale=0.8]{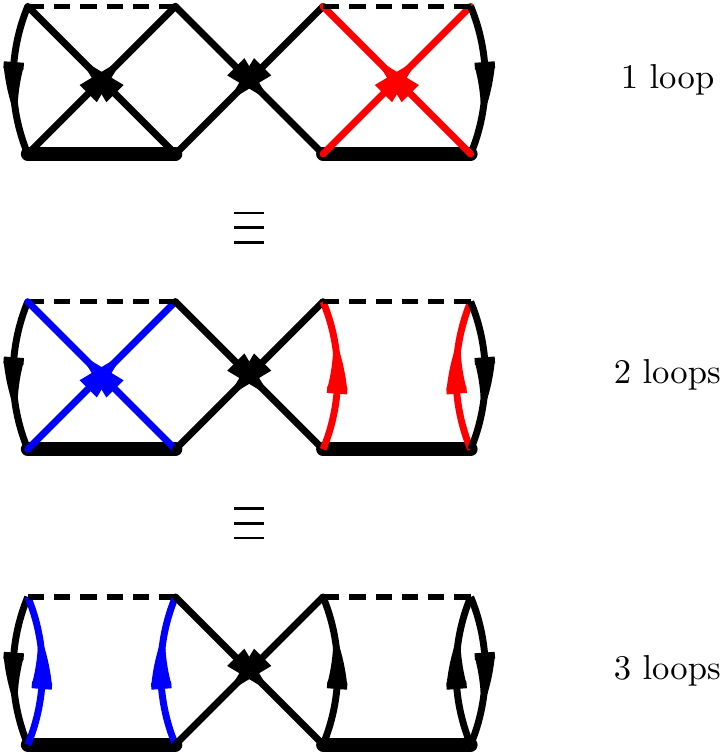}
\caption{\label{Antisym.fig} Using antisymmetry of tensors to alter the PHL structure of an examplatory Goldstone diagram. Red and blue lines are 
consecutively interchanged to reach a 1-loop, 2-loop and 3-loop diagram.}
\end{figure}
\end{itemize}

\subsection{Cost evaluation}
\label{CostEvaluation}

The key concept for the global optimization problem is the definition of a cost function $f$ evaluating the quality of individual diagrams. In general, low costs should correspond to a good diagrammatic quality and high costs to a 
diagram of poor quality. For Goldstone diagrams, several drawing conventions were defined, which are fulfilled by diagrams that 
are considered more elegant than others. These conventions should, in the first place, increase the readability of the individual diagrams and should somehow imitate human-drawn diagrams (i.e. should prefer symmetric diagrams or prefer 
diagrams without line crossings, etc.). 

\subsubsection{Static costs}
Whenever evaluating Goldstone diagrams, there are static cost factors appearing solely through the individual fragments of the diagram. Those static costs are summarized in table \ref{costs.tab} for 
different example diagrams. The individual cost values for each of the illustrated factors were chosen empirically.

\begin{table}
\tabcolsep=15pt
\caption{\label{costs.tab}Static costs of Goldstone diagrams during gene evaluation.}
\begin{tabular}{ c  Sc  c }
\toprule
Index & Example & Costs \\ \hline
(A) & \cincludegraphics[scale=0.5]{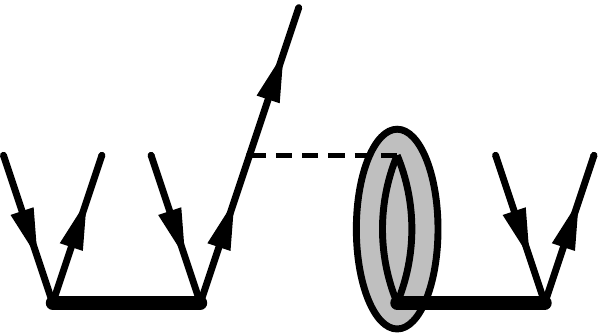}          & $-10$         \\ \\
(B) & \cincludegraphics[scale=0.5]{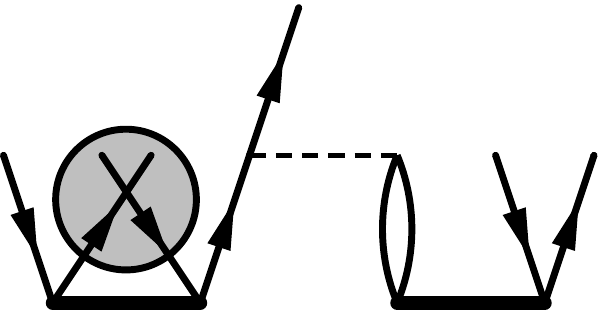}   & $+5$          \\  \\
(C) & \cincludegraphics[scale=0.5]{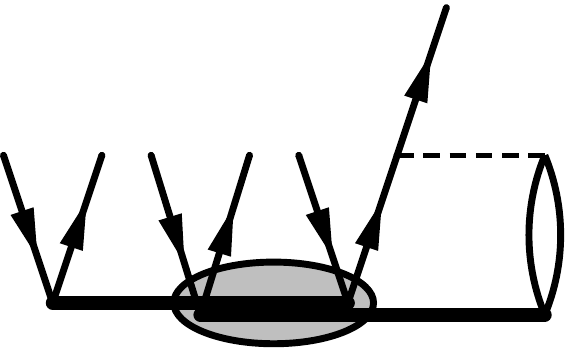}    & $+10$          \\  \\
(D) & \cincludegraphics[scale=0.5]{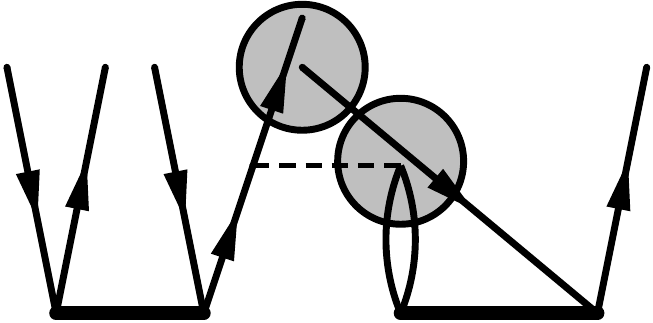} & $+2 \cdot 1.5$  \\  \\
(E) & \cincludegraphics[scale=0.5]{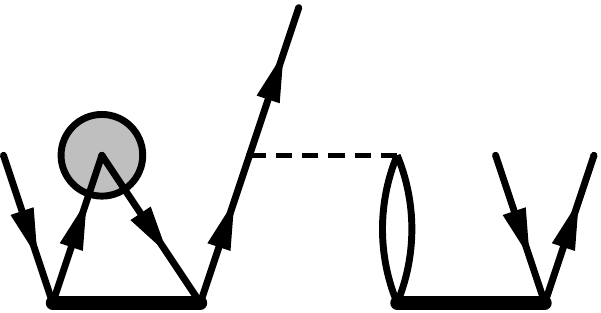} & $+100$          \\  \\
(F) & \cincludegraphics[scale=0.5]{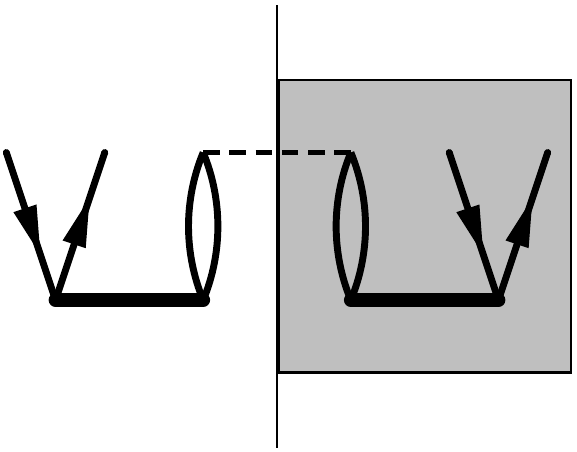} & $-1$          \\
\bottomrule
\end{tabular}
\end{table}

\begin{enumerate}
\item[(A)]{Since any form of loop is easily recognizable and is beneficial for a scaling analysis, the algorithm should always try to find as many loops as possible for the given diagram. Every loop therefore lowers the total costs by $10$\,units.}
\item[(B)]{Crossings of any kind on the other hand 
can be misleading and should be avoided. This is ensured by adding $5$\,units per PHL crossing to the total costs.}
\item[(C)]{Since interaction or cluster line crossings can only occur horizontally, 
a crossing of e.g. two $\hat{T}_2$ cluster lines may easily be misinterpreted by a $\hat{T}_4$ cluster line. Therefore, any interaction or cluster line crossing 
is rated worse than any PHL crossing by applying twice the cost of a PHL crossing.}
\item[(D)]{If two lines are close to each other, it is harder to distinguish between them. In such a case, $1.5$\,units are added to the total costs forcing all lines to lie within a certain minimal distance.}
\item[(E)]{Furthermore, $100$\,units are added whenever two vertices are arranged at the same 
position, which can also lead to irritating diagrams. In the illustrated case, two external vertices fall together forming a joined vertex, which could be wrongly recognized as a $\hat{T}_1^\dagger$ vertex.}
\item[(F)]{Finally, a symmetry check is executed evaluating the existence of a plane of reflection in the center of the diagram. If such a plane was found, one unit is subtracted from the total costs. }
\end{enumerate}

\subsubsection{Dynamic costs}
In addition to fixed costs, several dynamic (variable) costs are included in the implemented cost function. 
\begin{itemize}
\item[(I)] To ensure a reasonable line length, all particle/hole and interaction line lengths are calculated during the cost evaluation. 
For straight PHLs, a preferred line length of $\sqrt{\frac{10}{9}}$ was defined, which corresponds to the length of a line advancing one full unit (three thirds in the solution grid) upwards and one third of a unit sideways. 
For PHLs, which are part of a loop, the preferred line length corresponds to the difference in time-orders of the two interaction lines connected by the loop. Any absolute deviation of the calculated line lengths to the preferred ones 
is added to the total costs.

\item[(II)] Whenever two particle or two hole lines are connected to the same vertex, a straightness criterion is checked. If the found particle or hole lines are not collinear, the deviation in $x$-direction to collinearity is added to the total costs.
\end{itemize}


\section{Approach of solution: Genetic Algorithm}
\label{SolutionApproach}

The Goldstone diagram routing problem as defined in section \ref{OptimizationProblem} as a global minimization of the cost function $f$ represents a 
discrete high-dimensional optimization problem with exponentially increasing complexity with the number of vertices. This suggests the application 
of a genetic algorithm as outlined within the next subsections.

\subsection{General structure}
First introduced by Holland \cite{Holland_1975}, genetic algorithms resemble a particular field of algorithms that 
use techniques inspired by evolutionary biology. These techniques include selection, crossover (and therefore inheritance), 
mutation and death. Genetic algorithms are considered as efficient solution algorithms for high-dimensional, global optimization 
problems in a discrete solution space, \cite{Kumar_2010} which have been used in CC related research fields (see e.g. \cite{Engels_2011} used by \cite{Hanrath_2010}).

\begin{figure}
\includegraphics[scale=1]{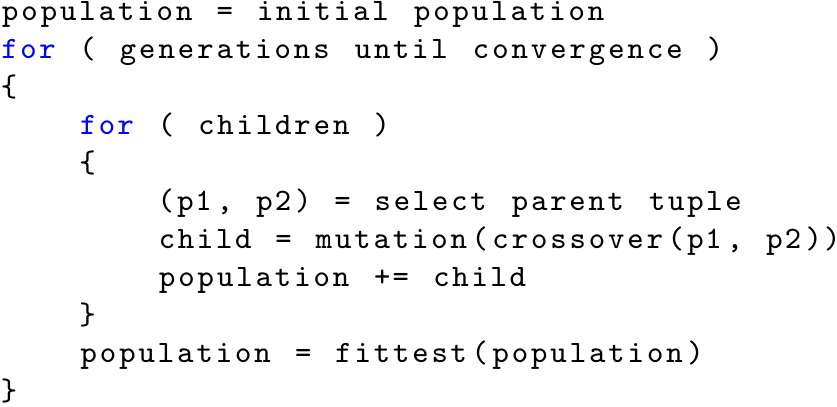}
\caption{\label{GS.fig}Sketch of the general structure of a genetic algorithm.}
\end{figure}

Figure \ref{GS.fig} shows a sketch of the steps involved a genetic algorithm as pseudo code.
Starting with the creation of random individuals, an initial guess of $n$ individuals is instantiated. In terms 
of genetics, these individuals are recognized as \textit{genes}, that make up the first \textit{population}.
Each individual gene is rated according 
to its quality with respect to the problem in question. Genes of higher quality may be rated with a lower cost (or a higher fitness) than genes of poorer quality.
Certain genes may then be selected as suitable mates, creating a new gene, which can be 
generated as a crossover of the mating genes. This step of creating genetic children is called \textit{crossover}. To ensure genetic diversity, every genetic child has 
a specific chance (or rate) to undergo a \textit{mutation} process, resembling a random change of its contents (its genetic 
material). Via these crossover and mutation processes, new genes are derived from already existing ones while keeping a controlled 
degree of random variation through the employed mutation rate. Especially for global optimization problems, a key concept of solution 
algorithms is the ability to converge across local minima into the desired global minimum. Even if the local minima are well spread within the 
whole solution space, the employed mutation process may grant this ability depending on the number of generations, the population size and the mutation rate.

After introducing all generated children into the population, the latter undergoes a \textit{selection} process, where the weakest genes are erased from it. This 
keeps the population size fixed to $n$ genes and resembles an important optimization step, since all inaccurate genes are removed and are therefore not allowed to pass 
their genetic material to possible children.
Finally, if no further evolutionary progress was made (e.g. no improvement was found) 
the algorithm terminates with the final result being found in the best gene of the final population. In every other case, the population will undergo all 
steps again, starting with the crossover of genes. 

Although the algorithm itself has a rather fixed structure, the exact formulation of crossover and mutation are strongly dependent on 
the underlying optimization problem and are crucial for the performance of the algorithm.

\subsection{Reproduction}

The reproductive process (crossover and mutation) resembles one of the most crucial steps of any genetic algorithm towards convergence. In particular for Goldstone diagrams, reproduction is neither trivial nor in some fashion canonically defined. As outlined in the next subsections, a genetic crossover (c.f. subsection \ref{Crossover}) based on diagram fragmentation was found to perform well while different mutational processes sorted into 
two mutational phases are employed (c.f. subsection \ref{Mutation}).

\subsubsection{Crossover}
\label{Crossover}

\begin{figure}
\includegraphics[width=\columnwidth]{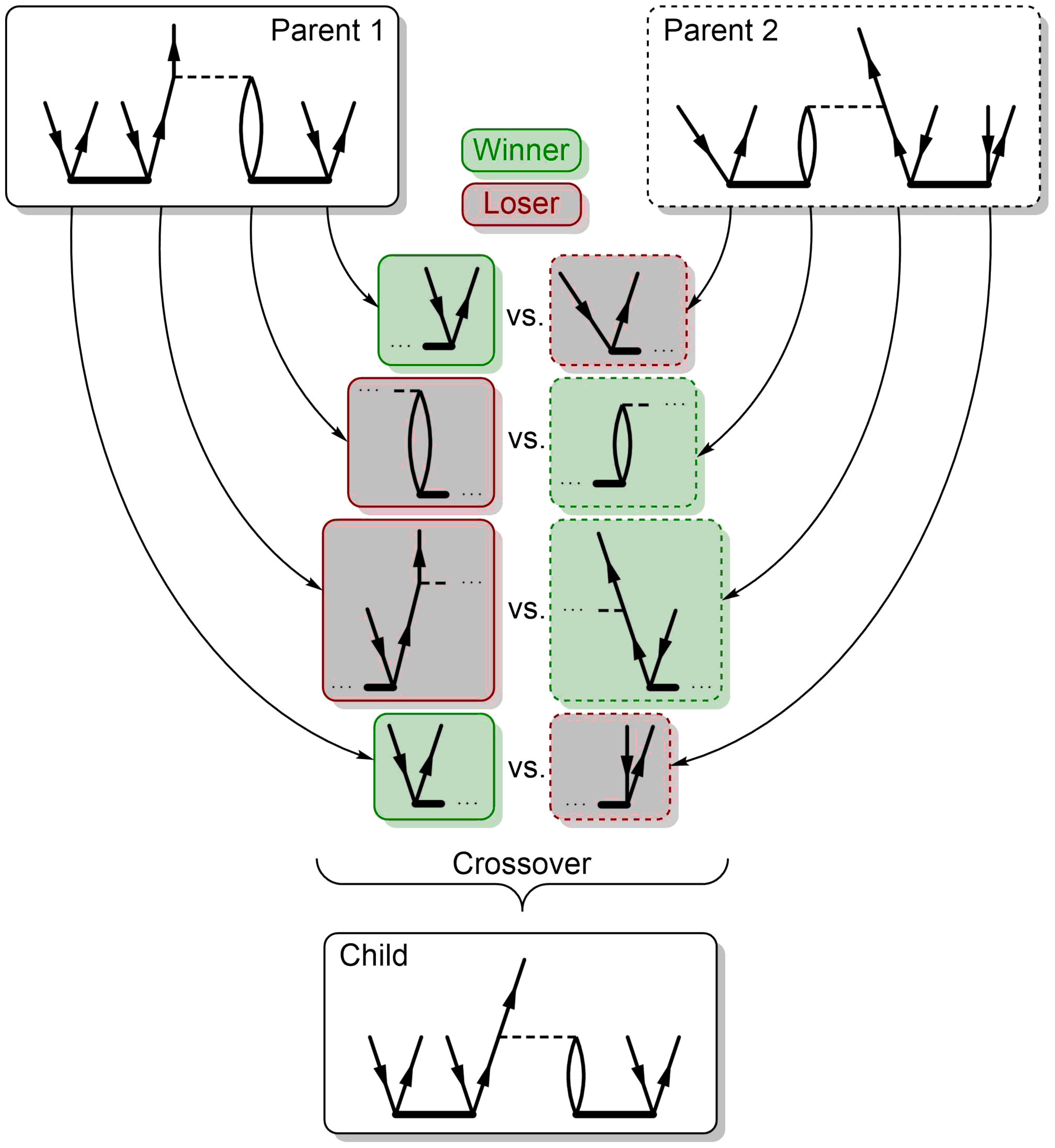}
\caption{\label{Reproduction.fig}Conceptual illustration of the fragment based crossover for Goldstone diagrams.}
\end{figure}

A definition of a reasonable genetic crossover of two Goldstone diagrams was found employing a fragment-based ansatz. The principle concept of this ansatz is illustrated in figure \ref{Reproduction.fig}.

In this ansatz, two parent Goldstone diagrams produce a child diagram via fragmentation, competition and reassignment. The individual parent diagrams are illustrated at the 
top of figure \ref{Reproduction.fig}. Both parents are fragmented 
into PHL segments 
(c.f. subsection \ref{PHLStructure}) by breaking the participating interaction or cluster lines. Those fragments are 
represented in solid and dashed frames, illustrating their parent affiliation. All corresponding 
fragments are then competing against 
each other, where the costs of the individual fragments are evaluated and compared.

For this fragmentation and cost evaluation, the parametrization of PHLs into individual segments is beneficial. Since all PHL segments are fragments per definition, no explicit fragmentation process
is needed. Iterating over all vertices contained in one PHL segment allows for an efficient fragment cost calculation, which is needed to compare the individual fragments later on. The fragment cost calculation itself 
is not different from a usual cost calculation as described in subsection \ref{CostEvaluation} checking all the proposed criteria only for the selected subset of all vertices (with a few obvious exceptions e.g. symmetry or length costs for interaction lines). 

In the final reproductive step, the best fragments are reassembled to result in the final child diagram. This is done by directly copying the $x$ and $y$ coordinates of the winning fragments 
to the child.

There are two major issues concerning the fragment based crossover ansatz:
\begin{enumerate}
\item[(i)]{The fragment based crossover is not always possible. For any successful reassignment, the fragments of both parents must be exactly equal to each other. Using the antisymmetry principle, the outgoing or incoming lines of any interaction or cluster line may be interchanged, thus creating different fragments. In any such cases, the two parent costs are directly compared to 
each other and the better parent diagram will be cloned.}
\item[(ii)]{The fragment's relative coordinates towards the other fragments are not taken into account. Therefore, many crossovers may lead to poor diagrams with equal vertices or many line crossings. With increasing generation however, 
only diagrams that are beneficially reproducible survive the algorithm. In fact, a direct comparison of the simple reassignment to a more sophisticated reassignment procedure revealed that much better results and a better efficiency were 
obtained when using the proposed simple approach. The collective intelligence of the genetic algorithm seems to reside in the selection process and not within a sophisticated crossover for this particular problem.}
\end{enumerate}

\subsubsection{Mutation}
\label{Mutation}

The mutation process assures the needed genetic diversity to inspect all possible areas of the solution space. It is responsible for a successful convergence of the algorithm to the global minimum. To ensure this diversity, the mutation process must be capable of reaching the global minimum by accident within a single mutation of any diagram. 
In particular, this includes the topology altering usage of antisymmetry (c.f. subsection \ref{DegreesOfFreedom}) throughout the whole algorithm. 

Additionally to an exponentially decreasing mutation rate, it was found benefical (faster converging) to define different mutational phases I and II, where different operations are carried out:

\quad \\
\textbf{Phase I: Large diversity via random absolute positioning} \\
Phase I features mutations of large diversity, where $x$- and $y$ coordinates of all vertices are randomly assigned to new values with the chance of the exponentially decreasing 
mutation rate.

\quad \\
\textbf{Phase II: Small diversity via differential updates} \\
Phase II features minor mutations, which employ differential updates to the vertex positioning. All possible mutations during phase II are illustrated in table \ref{mut.tab}. 

\begin{table}
\tabcolsep=9pt
\caption{\label{mut.tab}Possible Goldstone diagram mutations during phase II.}
\begin{tabular}{ c  Sc c Sc }
\toprule
(A) & \cincludegraphics[scale=0.4]{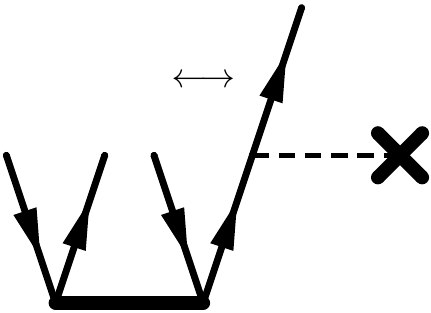} & $\leftrightarrow$ & \cincludegraphics[scale=0.4]{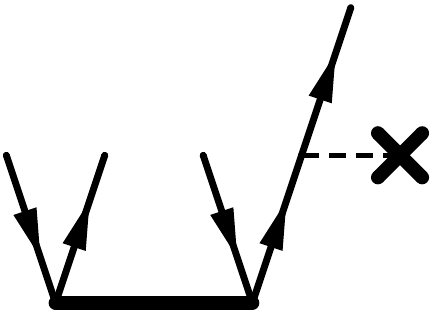} \\ 
(B) & \cincludegraphics[scale=0.4]{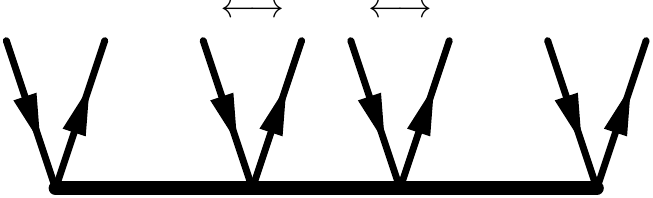}    & $\leftrightarrow$ & \cincludegraphics[scale=0.4]{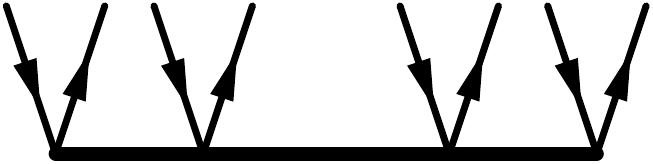} \\ 
(C) & \cincludegraphics[scale=0.4]{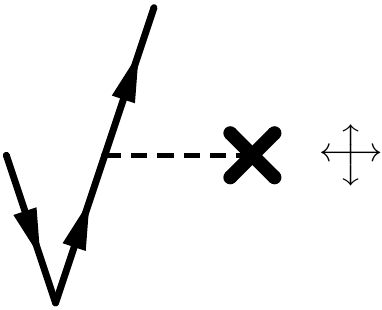}       & $\leftrightarrow$ & \cincludegraphics[scale=0.4]{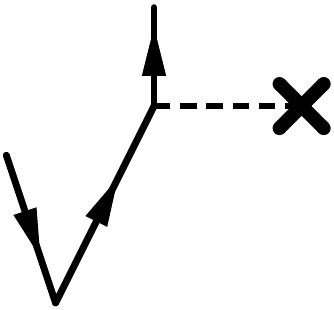} \\ 
(D) & \cincludegraphics[scale=0.4]{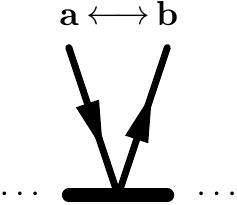}          & $\leftrightarrow$ & \cincludegraphics[scale=0.4]{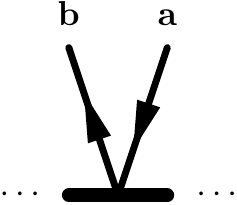} \\
\bottomrule
\end{tabular}
\end{table}
\begin{enumerate}
\item[(A)]{With the given mutation rate, certain PHL fragments are selected for a random movement in $x$ direction between $-1$ and $+1$\,units. All other vertices of all other fragments are kept constant. This may cure too long or too short interaction or cluster lines.}
\item[(B)]{A scissoring movement was employed to account for large interaction or cluster lines, where all vertices are not equally distributed. Unique pairs of PHL segments are chosen with the chance of the given mutation rate. These pairs are then moved towards or away from each other with $\{-1, \ldots, +1\}$\, (randomly chosen) units in $x$-direction.}
\item[(C)]{An interaction line shift was employed, moving all vertices associated with randomly chosen interaction or cluster lines by $\{-2, \ldots, +2\}$\, (randomly chosen) units in $x$-direction and by $\{-1, \dots, +1\}$\,units (if time order conserving) in $y$-direction.}
\item[(D)]{Finally, a flip of individual PHL segments was employed, were all vertices of a certain segment are mirrored with respect to a central plane of reflection. This mutation is particularly useful for diagrams containing $\hat{F}_N$
vertices, where the cross is oriented in an unfavorable direction.}
\end{enumerate}


\section{Convergence analysis}
\label{Algorithmanalysis}

In the following subsections, several analyses of the proposed algorithm are demonstrated. 

Subsection \ref{ConvergenceBehaviour} presents an an in-depth view on convergence and complexity of the genetic algorithm. An arbitrary example of large complexity
from the fixpoint iterative CCSD expansion is used. 

The reproductive process is analyzed in subsection \ref{ReproductionandMutationRatio} while timings of the algorithm 
are given in subsection \ref{Timings}.

\subsection{Complexity and convergence behavior}
\label{ConvergenceBehaviour}

\begin{figure*}
\includegraphics[width=0.8\textwidth]{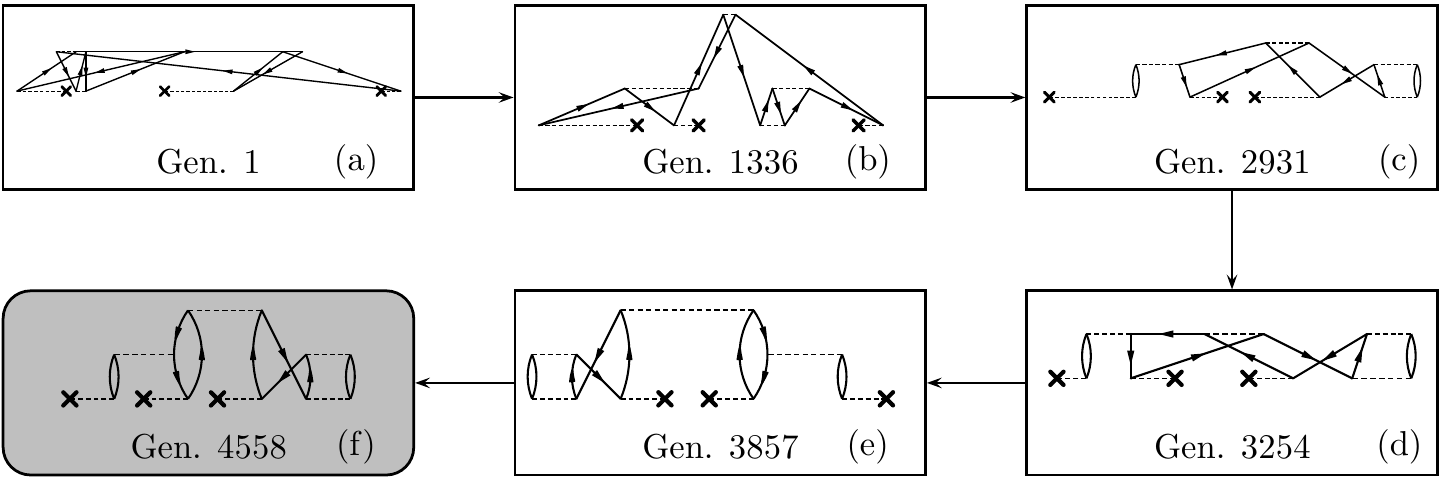}
\caption{\label{GOpt.fig}Goldstone diagram optimization route for an examplatory diagram of the fixpoint iterative CCSD expansion (constructed in fig. \ref{FixPointExample.fig})
showing specific diagrams of selected generations for which large improvements were found.}
\end{figure*}

In figure \ref{GOpt.fig}, the optimization route explicitly showing the leading diagrams throughout the optimization, is illustrated for the optimization of the diagram constructed in figure \ref{FixPointExample.fig}.

This particular optimization was conducted using the default settings of a population size of 40 genes times the number of vertices, which corresponds to $p = 560$ in the illustrated case. The convergence 
threshold of the algorithm was set to a cost difference from worst to best diagram of $10^{-6}$\,units.
The mutation rate was chosen to exponentially decay from $1.0$ to $0.1$ within 5600 generations ($g=10p$) (if not converged earlier). In case of no convergence after the 5600 generations the algorithm 
propagates with a constant mutation rate of $0.1$.
 As described 
in subsection \ref{Mutation}, two mutational phases were employed to reduce the number of generations needed to converge. 
Phase I (larger mutations) was applied from generation 1 to 2799 and phase II (weaker mutations) was applied from generation 2800 ($\frac{g}{2}$) to convergence. 

Figure \ref{GOpt.fig} (a) shows 
the best diagram of the initial population (created randomly) and (f) the converged diagram. Clearly, the initial diagram at generation 1 represents a poor solution guess with many 
PHL crossings. However, due to the used parametrization (c.f. section \ref{OptimizationProblem}), all diagrams automatically follow the correct virtual time ordering. This is even 
true for the first generation only consisting of randomly created diagrams. Within the large mutation range from generation 1 to 2799 to diagram (b), the algorithm
proceeds in avoiding the fundamental costs such as e.g. crossings or line vicinities. In the small mutation range afterwards through diagrams (c) to (f), the optimal routing is constructed,
applying only small shifts of single vertices or interaction lines.

The optimized diagram contains 14 vertices with degrees of freedom in their $x$-coordinates and three interaction line collections with unique time positions, which possess degrees of freedom in their $y$-coordinates. 
Therefore, the optimization problem contains $17$ positional degrees of freedom. The PHL structure for each $\hat{V}_N$ contributes two ($(2!)^2 - 2! = 2$) additional degrees of freedom. 

The minimal solution grid of thirds contains $19\times6$ unique vertex positions. Since there is no restriction 
that different vertices can not occupy the exact same position, the size of the possible solution space (neglecting time ordering) is exceedingly large with roughly
\begin{equation*}
19^{14}\cdot 6^3 \cdot 2^3 \approx 1.38\cdot 10^{21}.
\end{equation*}

Due to this large solution space, any brute force algorithm is impossible and the solution via a genetic algorithm in only 4558 iterations (generations) presents an efficient solution. 

\begin{figure}
\includegraphics[angle=-90, width=\columnwidth]{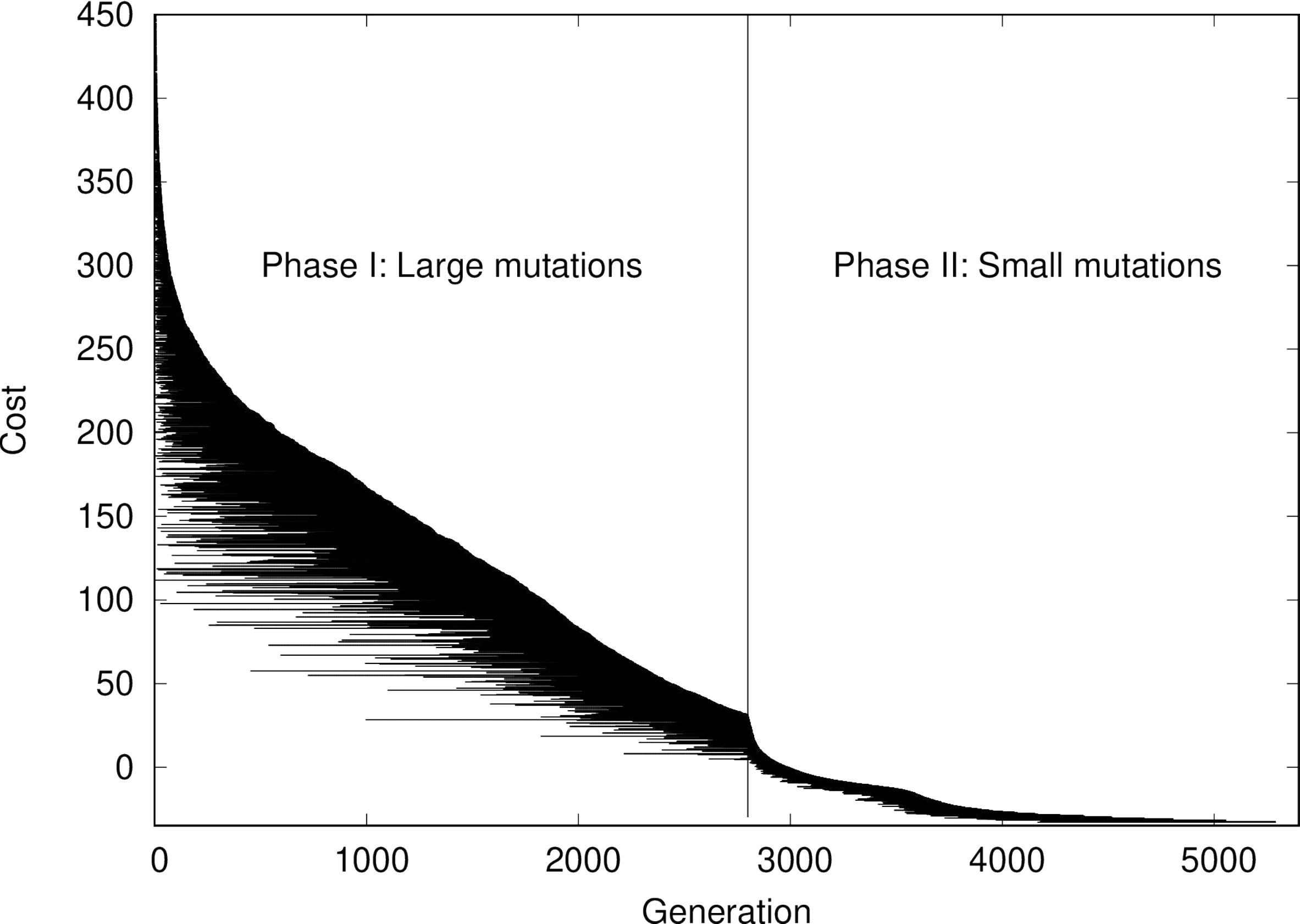}
\caption{\label{CostPlot_g.fig}Goldstone optimization course explicitly showing all costs of all diagrams in each generation.}
\end{figure}

To further analyze the convergence behavior, the optimization course was illustrated as a function of the individual diagram costs in figure \ref{CostPlot_g.fig}.

In this illustration, every gene of every population (in every generation) is described by a single point corresponding to its costs. Therefore, it is in principle possible to track how individual genes propagate through the genetic algorithm
from their first appearance to their potential death. 

During phase I, a fast decrease 
in costs within the first generations is recognizable. Subsequently, the costs decrease linearly up to the end of phase (I). The cost range from best to worst diagram is kept 
nearly constant. Therefore, the genetic diversity of all populations is large enough to ensure that the algorithm does not converge too early in any local minima. 

In phase II however, the cost range is drastically reduced.
The removal of phase II lead to exceedingly large generation numbers needed to converge the algorithm. 
As a compromise, the phase transition was implemented at $g=10n/2$, which produced optimal results for every tested diagram. 

\subsection{Crossover and mutation ratio}
\label{ReproductionandMutationRatio}
Children are created by directly combining the coordinates of the winning fragments from their parents. This may lead to diagrams which are 
worse compared to their parents. 
As stated in subsection \ref{Crossover}, the fragment-based crossover for Goldstone diagrams is not always capable of producing a genetically altered child. In any such cases, the algorithm will proceed by creating a 
clone of one of its parent diagrams. 

\begin{figure}
\includegraphics[angle=-90, width=\columnwidth]{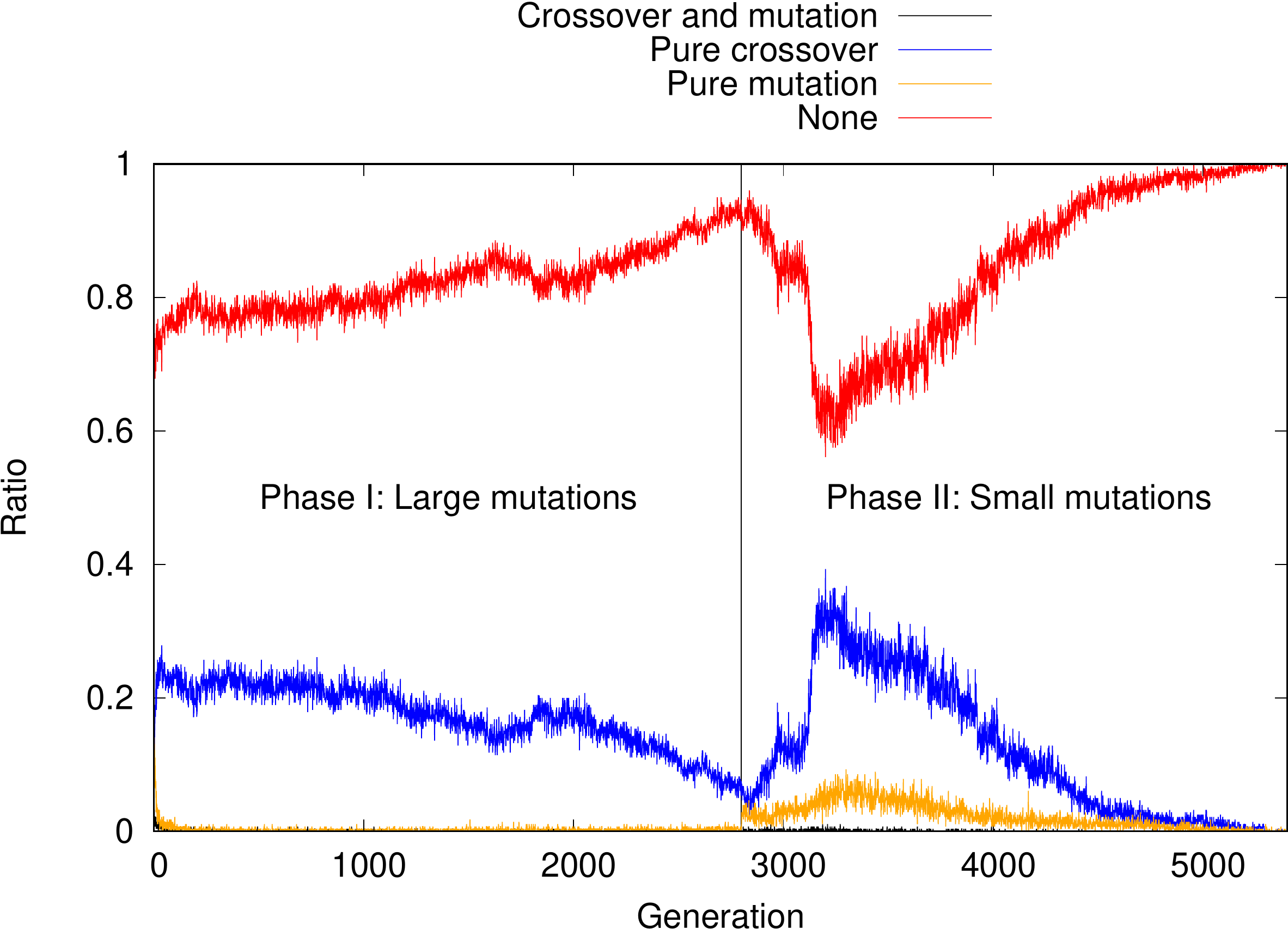}
\caption{\label{RepMutDist_g.fig}Ratio of improving crossovers and mutations during the Goldstone diagram optimization with respect to all $\frac{n}{2}$ children.}
\end{figure}

Using the same example of the diagram constructed in figure \ref{GOpt.fig},
the crossover and mutation processes were analyzed. 
This was done by calculating the number of diagrams improved by crossover (with respect to both their parents) and by mutation, purely by crossover, purely by mutation and by none of both. The percental ratios from all created 
children during each generation are illustrated in figure \ref{RepMutDist_g.fig}.

Clearly, pure crossover ($20\%$) dominates over combined crossover and mutation and pure mutation during the beginning of the algorithm. This effect results from the large mutation rate and range 
at the beginning of the algorithm.
While mutation keeps a large genetic diversity in the algorithm (by producing diagrams in every area of the solution space), pure crossover can produce an improved child in nearly every fifth case. Towards the end 
of phase I, pure crossover slowly decreases below 10\% until phase II is entered. 

Due to the smaller mutation range in phase II, improvement by pure crossover starts increasing instantly while improvement by pure mutation enters with roughly 5\% to 10\% of improved children. 
While the improvement by pure mutation stays low around the 5\% mark, the improvement by pure crossover decreases again until the final converged diagram is constructed at generation 4558. After this point pure crossover and pure mutation 
tend towards 0\% since there is no additional cost gain from altering the latter diagram.

A similar picture may be obtained for different diagrams of comparable complexity and it is therefore possible to conclude 
that the fragment-based crossover works.

\subsection{Timings}
\label{Timings}
The computing time of the genetic algorithm was statistically analyzed for a large collection of diagram optimizations. These include different CC approaches including different Hamiltonians (normal, three-particle interrelating and $\hat H_N^2$) as well as 
variational and similarity transformed CC approaches. The time ordered topologies were obtained by the term generation engine \enquote{sqdiag} \cite{sqdiag1, sqdiag2} and the optimizations 
were conducted on a typical office machine using a single core of the Intel Core i3-8100 CPU with 3.60\,GHz.

\begin{figure}
\includegraphics[width=\columnwidth]{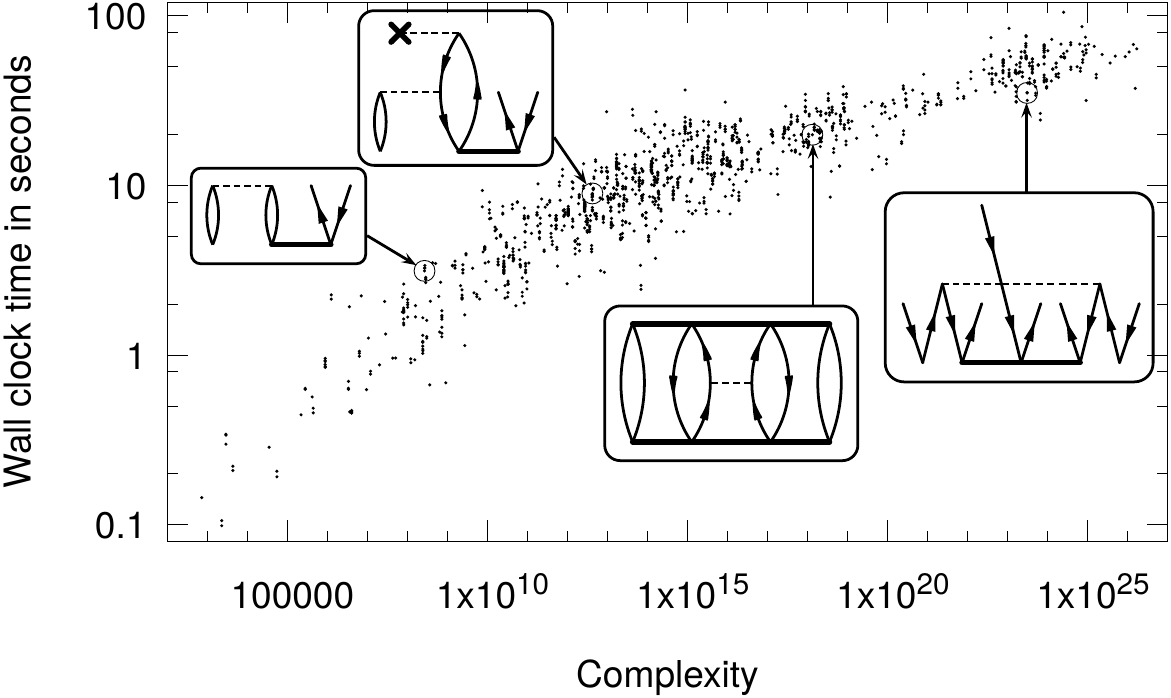}
\caption{\label{ComplexityVSTime.fig}Wall clock time with respect to diagrammatic complexity for a large collection of different diagrams (dots) with four representatives explicitly shown.}
\end{figure}

In figure \ref{ComplexityVSTime.fig}, the individual diagrammatic complexities are plotted against the wall clock computing time (on a single computing core) needed to achieve convergence.
The complexities were calculated accordingly to subsection \ref{ConvergenceBehaviour}.

Clearly, an increased complexity of the solution space in general leads to higher computing times.  A large broadening effect is recognizable, where different optimizations
of comparable complexity may take different computing times (up to two orders of magnitude). This may result from various starting configurations of the initial population, which may take different numbers of generations to converge.  

\begin{figure}
\includegraphics[width=\columnwidth]{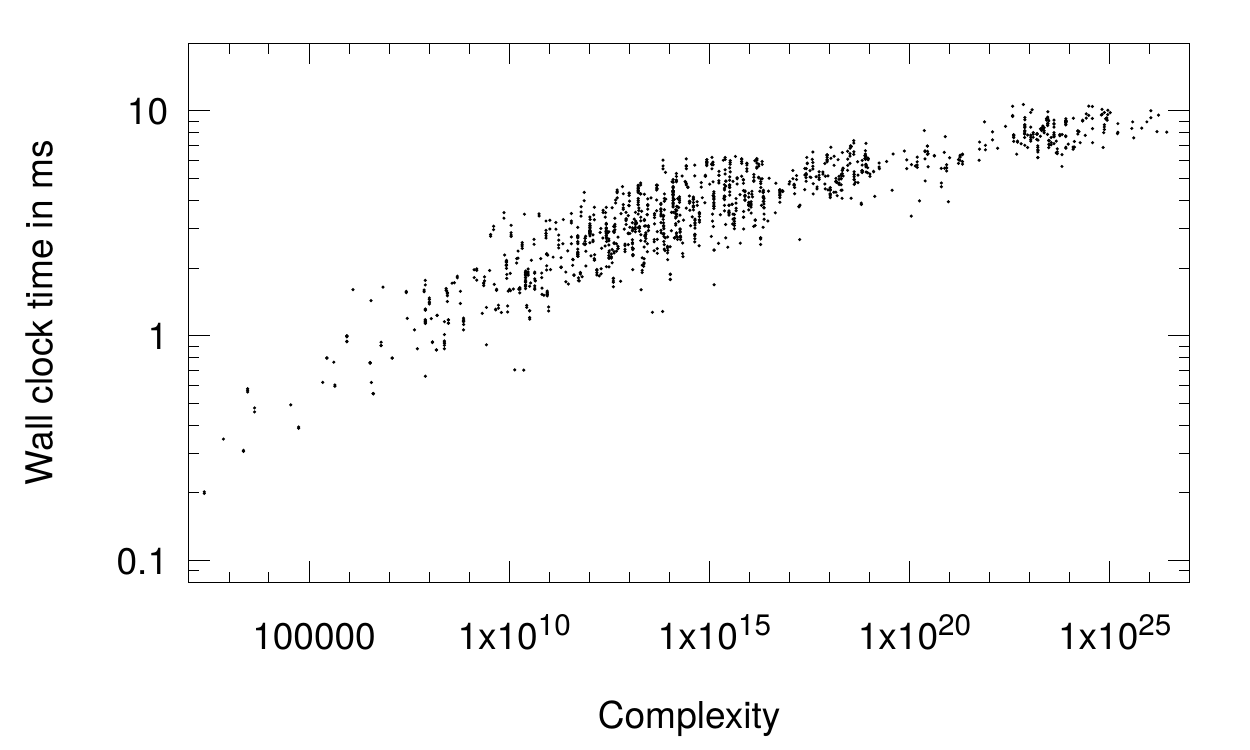}
\caption{\label{TimingTime.fig}Wall clock time per generation in ms with respect to diagrammatic complexity for a large collection of different diagrams.}
\end{figure}

Due to these different generation numbers, any direct relation of diagrammatic complexities to real computation times is biased. Therefore, the average wall clock time for one generation was calculated and illustrated 
with respect to the diagrammatic complexity (c.f. figure \ref{TimingTime.fig}).

In general, wall clock times of under 1\,ms to 10\,ms per generation were found. Again, 
a slight broadening effect is visible, which appears in a small range of under 10\,ms.

Overall, sufficiently small wall clock times were found such that no conducted optimization took longer than 100 seconds. It is important to state that up to this point no thorough code optimization was conducted. So far the emphasis was on functionality of the algorithm rather than efficiency. 
Additional code optimizations would presumably lower the total computation time significantly.

\section{Application}
\label{application}

Several Goldstone diagram optimizations were conducted from different CC approaches to check the functionality of the developed algorithm. This includes example diagrams from
\begin{itemize}
\item{the standard similarity transformed CCSDTQ ansatz (up to quadruple projections),}
\item{the variational CC ansatz for CCSDTQ (only linear terms $e^{\hat{T}} \approx 1 + \hat{T}$),}
\item{the variational CC ansatz for CCSD (linear and quadratic terms $e^{\hat{T}} \approx 1 + \hat{T} + \hat{T}^2$),}
\item{the similarity transformed CCSD ansatz employing $\hat{H}_N^2$ (up to single projections) and}
\item{the similarity transformed CCSDT ansatz for $\hat{H}_{N3}$ (up to triple projections).}
\end{itemize}

The time-ordered diagram topologies were obtained from the term generation engine \enquote{sqdiag} \cite{sqdiag1, sqdiag2}.
To illustrate the quality of the obtained diagram layouts, a selected collection is given in the supporting information.
All diagram optimizations were 
conducted multiple times for different initial populations. All of those optimizations, found the same or qualitatively equal solutions even for initial populations of poor quality. To discuss benefits 
and potential flaws of the developed algorithm, a small collection of diagrams with mixed complexity from all conducted optimizations is illustrated in table \ref{Application.tab}.

\begin{table}
\tabcolsep=12pt
\caption{\label{Application.tab}Collection of optimized Goldstone diagrams for different CC approaches. For each contraction pattern only one participating diagram is shown.}
\begin{tabular}{ c  c  Sc }
\toprule
Bra      & Ket & Diagram \\ \hline
$\bra{\Psi_{IJK}^{ABC}}$ & 
$
\contraction[1ex]{}{\hat{V}_N}{}{\hat{T}_1}
\contraction[1.5ex]{}{\hat{V}_N}{\hat{T}_1}{\hat{T}_2}
\contraction[2ex]{}{\hat{V}_N}{\hat{T}_1}{\hat{T}_2}
\contraction[2.5ex]{}{\hat{V}_N}{\hat{T}_1\hat{T}_2}{\hat{T}_2}
$
$\hat{V}_N\hat{T}_1\hat{T}_2\hat{T}_2\Psi_0\rangle$\footnotemark[1]\rule{0pt}{6ex} & 
\cincludegraphics[scale=0.3]{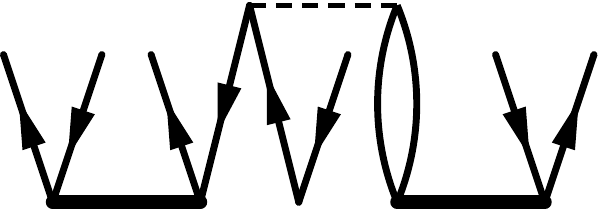}\\

$\bra{\Psi_{IJKL}^{ABCD}}$ & 
$
\contraction[1.0ex]{}{\hat{V}_N}{}{\hat{T}_2}
\contraction[1.5ex]{}{\hat{V}_N}{\hat{T}_2}{\hat{T}_2}
$
$\hat{V}_N\hat{T}_2\hat{T}_2\Psi_0\rangle$\footnotemark[1] \rule{0pt}{5ex}& 
\cincludegraphics[scale=0.3]{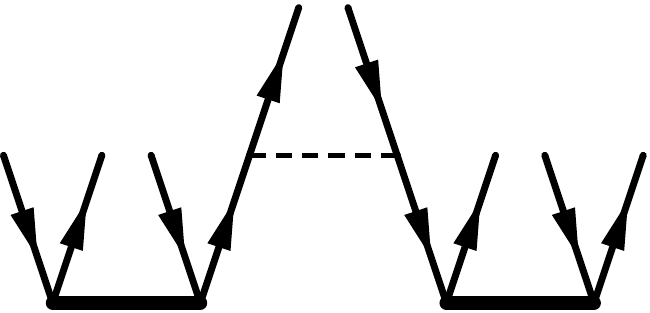}\\
\\\\
$\bra{\Psi_0}$ & 
$
\contraction[1.0ex]{\hat{T}^{\dagger}_2}{\hat{T}^{\dagger}_2}{}{\hat{F}_N}
\contraction[1.5ex]{\hat{T}^{\dagger}_2}{\hat{T}^{\dagger}_2}{}{\hat{F}_N}
\contraction[2.0ex]{\hat{T}^{\dagger}_2}{\hat{T}^{\dagger}_2}{\hat{F}_N}{\hat{T}_1}
\contraction[2.5ex]{}{\hat{T}^{\dagger}_2}{\hat{T}^{\dagger}_2 \hat{F}_N }{\hat{T}_1}
\contraction[3.0ex]{\hat{T}^{\dagger}_2}{\hat{T}^{\dagger}_2}{\hat{F}_N \hat{T}_1}{\hat{T}_2}
\contraction[3.5ex]{}{\hat{T}^{\dagger}_2}{\hat{T}^{\dagger}_2 \hat{F}_N \hat{T}_1}{\hat{T}_2}
\contraction[4.0ex]{}{\hat{T}^{\dagger}_2}{\hat{T}^{\dagger}_2 \hat{F}_N \hat{T}_1}{\hat{T}_2}
\contraction[4.5ex]{}{\hat{T}^{\dagger}_2}{\hat{T}^{\dagger}_2 \hat{F}_N \hat{T}_1}{\hat{T}_2}
$
$\hat{T}^{\dagger}_2 \hat{T}^{\dagger}_2 \hat{F}_N \hat{T}_1 \hat{T}_2 \Psi_0\rangle$\footnotemark[2] \rule{0pt}{8ex}& 
\cincludegraphics[scale=0.3]{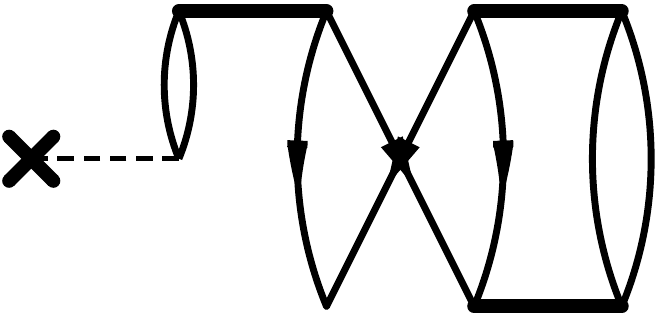}\\ 

$\bra{\Psi_0}$ & 
$
\contraction[1.0ex]{\hat{T}^{\dagger}_2 \hat{T}^{\dagger}_2}{\hat{V}_N}{}{\hat{T}_2}
\contraction[1.5ex]{\hat{T}^{\dagger}_2 \hat{T}^{\dagger}_2}{\hat{V}_N}{}{\hat{T}_2}
\contraction[2.0ex]{\hat{T}^{\dagger}_2}{\hat{T}^{\dagger}_2}{\hspace{-0.5pt}}{\hat{V}_N}
\contraction[2.5ex]{}{\hat{T}^{\dagger}_2}{\hat{T}^{\dagger}_2\hspace{-0.5pt}}{\hat{V}_N}
\contraction[3.0ex]{\hat{T}^{\dagger}_2}{\hat{T}^{\dagger}_2}{\hat{V}_N}{\hat{T}_2}
\contraction[3.5ex]{\hat{T}^{\dagger}_2}{\hat{T}^{\dagger}_2}{\hat{V}_N}{\hat{T}_2}
\contraction[4.0ex]{\hat{T}^{\dagger}_2}{\hat{T}^{\dagger}_2}{\hat{V}_N \hat{T}_2}{\hat{T}_2}
\contraction[4.5ex]{}{\hat{T}^{\dagger}_2}{\hat{T}^{\dagger}_2 \hat{V}_N \hat{T}_2}{\hat{T}_2}
\contraction[5.0ex]{}{\hat{T}^{\dagger}_2}{\hat{T}^{\dagger}_2 \hat{V}_N \hat{T}_2}{\hat{T}_2}
\contraction[5.5ex]{}{\hat{T}^{\dagger}_2}{\hat{T}^{\dagger}_2 \hat{V}_N \hat{T}_2}{\hat{T}_2}
$
$\hat{T}^{\dagger}_2 \hat{T}^{\dagger}_2 \hat{V}_N \hat{T}_2 \hat{T}_2 \Psi_0\rangle$\footnotemark[2]\rule{0pt}{9ex} & 
\cincludegraphics[scale=0.3]{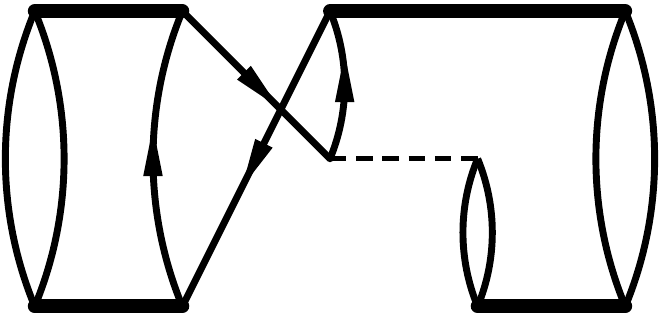}\\

\\\\

$\bra{\Psi_0}$ & 
$
\contraction[1.0ex]{}{\hat{V}_N}{}{\hat{F}_N}
\contraction[1.5ex]{}{\hat{V}_N}{\hat{F}_N}{\hat{T}_1}
\contraction[2.0ex]{}{\hat{V}_N}{\hat{F}_N}{\hat{T}_1}
$
$\hat{V}_N\hat{F}_N\hat{T}_1\Psi_0\rangle$\footnotemark[3] \rule{0pt}{5.5ex}& 
\cincludegraphics[scale=0.3]{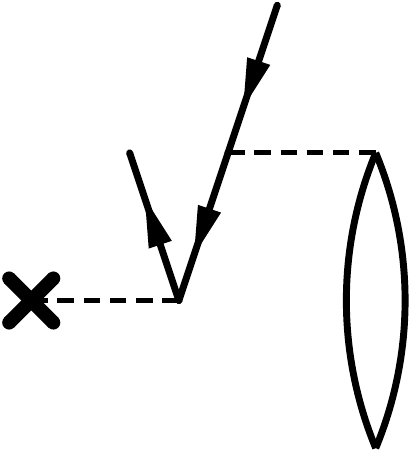}\\

$\bra{\Psi_0}$ & 
$
\contraction[1.0ex]{}{\hat{V}_N}{\hspace{-0.5pt}}{\hat{V}_N}
\contraction[1.5ex]{}{\hat{V}_N}{\hspace{-0.5pt}}{\hat{V}_N}
\contraction[2.0ex]{\hat{V}_N}{\hat{V}_N}{}{\hat{T}_1}
\contraction[2.5ex]{\hat{V}_N}{\hat{V}_N}{\hat{T}_1}{\hat{T}_1}
\contraction[3.0ex]{}{\hat{V}_N}{\hat{V}_N \hat{T}_1 \hat{T}_1}{\hat{T}_1}
\contraction[3.5ex]{}{\hat{V}_N}{\hat{V}_N \hat{T}_1 \hat{T}_1}{\hat{T}_1}
$
$\hat{V}_N \hat{V}_N \hat{T}_1 \hat{T}_1 \hat{T}_1 \Psi_0\rangle$\footnotemark[3] \rule{0pt}{7ex}& 
\cincludegraphics[scale=0.3]{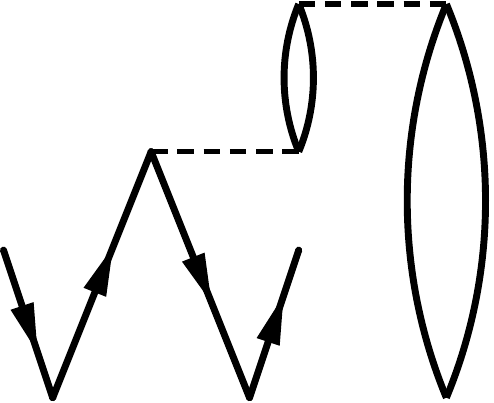}\\

\\\\

$\bra{\Psi_{IJ}^{AB}}$ & 
$
\contraction[1.0ex]{}{\hat{W}_N}{}{\hat{T}_2}
\contraction[1.5ex]{}{\hat{W}_N}{}{\hat{T}_2}
\contraction[2.0ex]{}{\hat{W}_N}{}{\hat{T}_2}
$
$\hat{W}_N\hat{T}_2\Psi_0\rangle$\footnotemark[4] \rule{0pt}{5.5ex}& 
\cincludegraphics[scale=0.3]{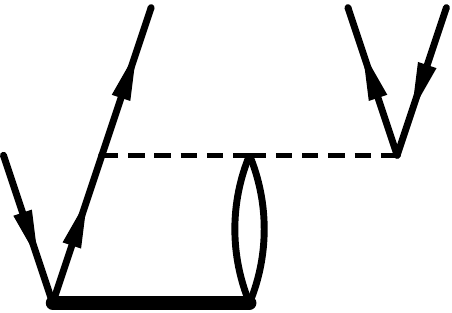}\\

$\bra{\Psi_{IJK}^{ABC}}$ & 
$
\contraction[1.0ex]{}{\hat{W}_N}{}{\hat{T}_1}
\contraction[1.5ex]{}{\hat{W}_N}{\hat{T}_1}{\hat{T}_2}
\contraction[2.0ex]{}{\hat{W}_N}{\hat{T}_1}{\hat{T}_2}
$
$\hat{W}_N\hat{T}_1\hat{T}_2\Psi_0\rangle$\footnotemark[4] \rule{0pt}{5.5ex}& 
\cincludegraphics[scale=0.3]{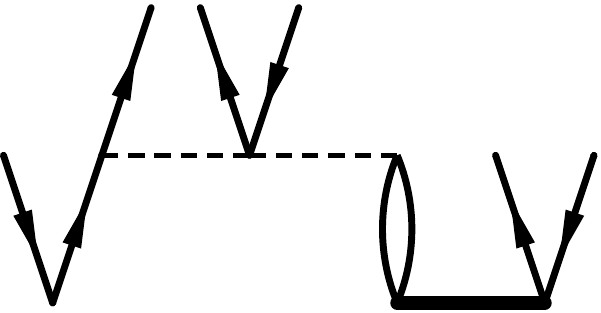}
\\
\bottomrule
\\[-23pt]
\footnotetext[1]{standard CC}
\footnotetext[2]{expectation value/variational CC}
\footnotetext[3]{variance calculation}
\footnotetext[4]{three body interactions}
\end{tabular}
\end{table}

All diagrams were successfully optimized to possess the maximum amount of loops, as few line crossings as possible (particle/hole, interaction and cluster) and no too close PHLs increasing the readability of the latter.
Furthermore, more complex loops containing three or more vertices are correctly optimized and illustrated. Even the 4-loop case always containing a PHL 
crossing (e.g. in the variational CC ansatz) is illustrated with circular PHLs at the outer sides signalizing its loop status.
Additionally, symmetric diagrams are preferred (if there is no additional loop by breaking symmetry).
We conclude that the developed algorithm is capable of producing essentially optimal Goldstone diagrams.

\begin{figure*}
\includegraphics[width=0.8\textwidth]{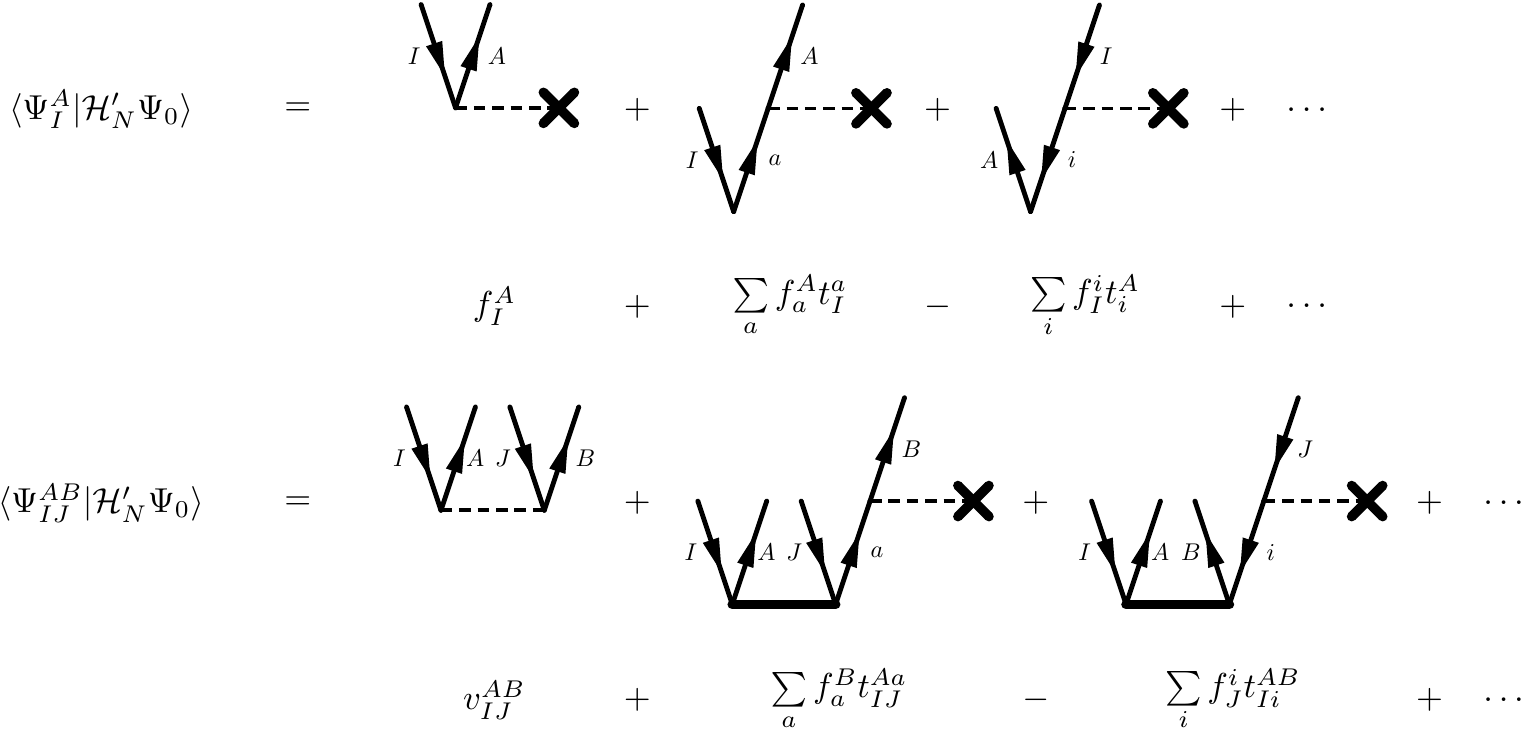}
\caption{\label{SDAmps.fig}Diagrammatic and algebraic amplitude CC equations for standard similarity transformed CCSD.}
\end{figure*}

\subsection{CC fixpoint iterations}
\label{CCSDfixpointiterations}

To further investigate the functionality of the developed algorithm, the CC fixpoint iterations were recreated in a purely diagrammatic representation. These present a good testing scenario  
because new diagrams with rapidly increasing complexity are generated only from the standard CC diagrams by diagrammatic substitutions (see e.g. \cite{Crawford_Book}). Therefore, no further input from the user is required 
once all diagram topologies are read in and the program is started. The proposed algorithm is capable of arbitrary Goldstone diagram substitutions (arbitrary CC truncation). 
Here, the results for CCSD are discussed in detail.

Considering the CCSD equations of the zeroth iteration, all amplitudes are set to zero with
\begin{equation}
t_i^a = t_{ij}^{ab} = 0 \qquad\forall_{ijab}.
\end{equation}
This of course directly results in a zeroth iteration correlation energy $E_\text{corr}^{(0)}$ of zero, since in every CCSD energy diagram either T$_1$ or T$_2$ amplitudes are present: 
\begin{align*}
\begin{tabular}{c c Sc c Sc c Sc}
$E_\text{corr}^\text{(CC)}$ & $=$ & $\vcenter{\hbox{\includegraphics[scale=0.4]{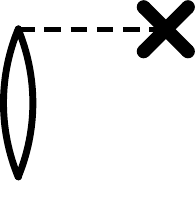}}}$
& $+$ & $\vcenter{\hbox{\includegraphics[scale=0.4]{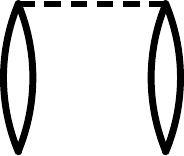}}}$
& $+$ & $\vcenter{\hbox{\includegraphics[scale=0.4]{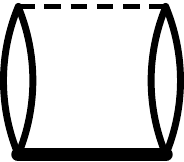}}}$
\end{tabular}
\end{align*}

The new first order T$_1$ and T$_2$ amplitudes are obtained from the singles and doubles projections (c.f. figure \ref{SDAmps.fig}). Since all amplitudes are zero, it is 
\begin{align*}
t_I^a &= t_i^A = t_I^A \qquad \forall_{ia} \\
t_{IJ}^{Aa} &= t_{Ii}^{AB} = t_{IJ}^{AB} \qquad \forall_{ia}
\end{align*}
and therefore the first three terms of the singles and doubles equations as illustrated in figure \ref{SDAmps.fig} simplify to 
\begin{align*}
\braket{\Psi_I^A|\bar{H}\Psi_0} &= f_I^A + \underbrace{\sum_{ia}(f_a^A - f_I^i)}_{-D_1}t_I^A = 0 \\
\braket{\Psi_{IJ}^{AB}|\bar{H}\Psi_0} &= v_{IJ}^{AB} + \underbrace{\sum_{ia}(f_a^B - f_J^i)}_{-D_2}t_{IJ}^{AB} = 0 \;.
\end{align*}

\begin{figure*}
\includegraphics[width=\textwidth]{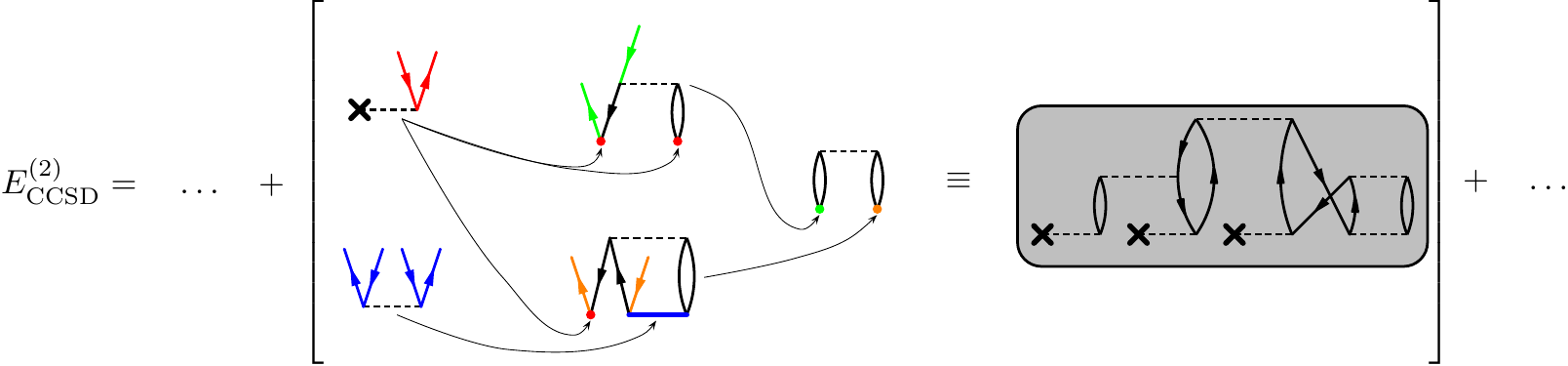}
\caption{\label{FixPointExample.fig} Substitution route of an examplatory Goldstone diagram from the second iteration of the fixpoint iterative CCSD expansion.}
\end{figure*}

Algebraic manipulation then leads to the first iteration T$_1$ and T$_2$ amplitudes consisting of the only diagrams not containing any amplitudes (purely external diagrams):

\begin{center}
\begin{tabular}{c c c c Sc}
$D_1t_I^A$ & $=$ & $f_I^A$ & $=$ & \cincludegraphics[scale=0.4]{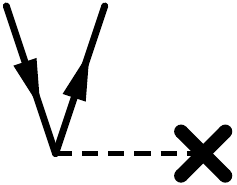} \\
$D_2t_{IJ}^{AB}$ & $=$ & $v_{IJ}^{AB}$ & $=$ & \cincludegraphics[scale=0.4]{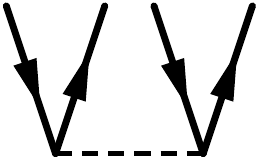}
\end{tabular}
\end{center}

For post-canonical HF CCSD the first iteration T$_1$ amplitudes are zero due to the diagonalized Fock matrix, for which
\begin{equation*}
f_I^A = 0 \qquad \forall_{I\in\mathbb{O}, A\in\mathbb{V}} \text{ with } \mathbb{O}\cap\mathbb{V} = \emptyset
\end{equation*}
holds.

For the sake of the algorithm testing however, post-non-canonical HF CCSD diagrams were considered to gain an even more rapidly increasing diagram complexity. 
In figure \ref{FixPointExample.fig}, one particular example of the second iteration CCSD energy expression is constructed. 

Table \ref{FixPoint.tab} shows a small collection of optimized Goldstone diagrams of different complexity from the second iteration correlation energy. All diagrams occuring
in zeroth, first and second iteration (energy, singles and doubles) are illustrated in the supporting information. In perturbation theory, it is customary to
keep track of the energy denominators (e.g. $D_1$ and $D_2$) by the inclusion of horizontal lines in the corresponding Goldstone diagrams. Every PHL crossed by these horizontal lines contributes to the energy denominator. \cite{Harris_Book} For the particular fixpoint iterative procedure presented here, these lines are not explicitly shown. It is also important to note that the usual rules 
to translate perturbation theory diagrams to their algebraic expressions (including their energy denominator) do not apply to the diagrams shown here since only those PHLs explicitly part 
of the substitution contribute to either one $D_1$ or one $D_2$ denominator. To illustrate the fixpoint substitution, one interaction line (from the original leftmost $\hat{F}_N$ or $\hat{V}_N$) is drawn at the highest possible 
relative position in time.

Clearly, all illustrated diagrams (A to F) as well as all fixpoint iterative diagrams in the supportig information are reasonably optimized. Every single one of them
is easily readable, analyzable and possesses the physically correct time ordering. The algorithm is capable of the optimization of several 
loop constructs. These include the \glqq{l}adder\grqq{} 3-loop (c.f. diagram E), the crossed 4-loop (c.f. diagrams A, D, F) as well as the direct 2-loop
(c.f. diagrams B, C, D, E, F) or combinations of all. Furthermore, antisymmetry is benefically used to increase the number of loops of all diagrams as e.g. clearly 
visible for diagram C. 

\begin{table}
\tabcolsep=12pt
\caption{\label{FixPoint.tab}Collection of optimized Goldstone diagrams of the second iteration correlation energy in a fixpoint iterative post-non-canonical HF CCSD.}
\begin{tabular}{ c  c  Sc }
\toprule
Index      & Contraction & Diagram \\ \hline
A &
$
\contraction[1.0ex]{\hat{F}_N}{\hat{F}_N}{}{\hat{F}_N}
\contraction[1.5ex]{\hat{F}_N}{\hat{F}_N}{\hat{F}_N}{\hat{F}_N}
\contraction[2.0ex]{}{\hat{F}_N}{\hat{F}_N}{\hat{F}_N}
\contraction[2.5ex]{}{\hat{F}_N}{\hat{F}_N\hat{F}_N}{\hat{F}_N}
\hat{F}_N\hat{F}_N\hat{F}_N\hat{F}_N
$ & \cincludegraphics[scale=0.4]{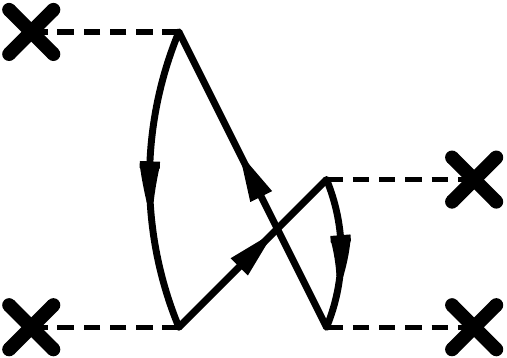} \\
B &
$
\contraction[1.0ex]{\hat{F}_N}{\hat{F}_N}{}{\hat{V}_N}
\contraction[1.5ex]{\hat{F}_N}{\hat{F}_N}{}{\hat{V}_N}
\contraction[2.0ex]{}{\hat{F}_N}{\hat{F}_N}{\hat{V}_N}
\contraction[2.5ex]{}{\hat{F}_N}{\hat{F}_N}{\hat{V}_N}
\hat{F}_N\hat{F}_N\hat{V}_N
$ & \cincludegraphics[scale=0.4]{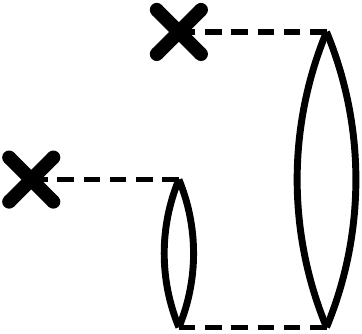} \\&&\\&&\\
C &
$
\contraction[1.0ex]{}{\hat{V}_N}{}{\hat{F}_N}
\contraction[1.5ex]{}{\hat{V}_N}{}{\hat{F}_N}
\contraction[1.0ex]{\hat{V}_N\hat{F}_N}{\hat{F}_N}{}{\hat{V}_N}
\contraction[1.5ex]{\hat{V}_N\hat{F}_N}{\hat{F}_N}{}{\hat{V}_N}
\contraction[2.0ex]{}{\hat{V}_N}{\hat{F}_N\hat{F}_N}{\hat{V}_N}
\contraction[2.5ex]{}{\hat{V}_N}{\hat{F}_N\hat{F}_N}{\hat{V}_N}
\hat{V}_N\hat{F}_N\hat{F}_N\hat{V}_N
$ & \cincludegraphics[scale=0.4]{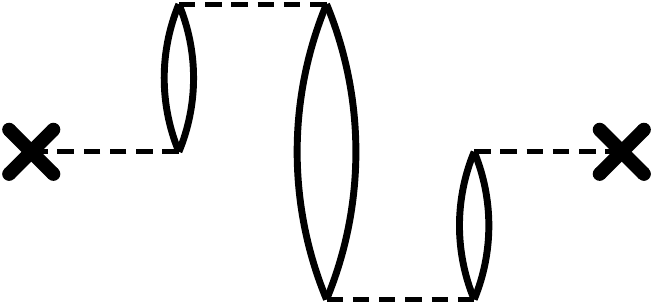} \\
D &
$
\contraction[1.0ex]{}{\hat{V}_N}{}{\hat{F}_N}
\contraction[1.5ex]{}{\hat{V}_N}{}{\hat{F}_N}
\contraction[3.0ex]{}{\hat{V}_N}{\hat{F}_N\hat{V}_N}{\hat{V}_N}
\contraction[3.5ex]{}{\hat{V}_N}{\hat{F}_N\hat{V}_N\hat{V}_N}{\hat{F}_N}
\contraction[1.0ex]{\hat{V}_N\hat{F}_N}{\hat{V}_N}{}{\hat{V}_N}
\contraction[1.5ex]{\hat{V}_N\hat{F}_N}{\hat{V}_N}{}{\hat{V}_N}
\contraction[2.0ex]{\hat{V}_N\hat{F}_N}{\hat{V}_N}{}{\hat{V}_N}
\contraction[2.5ex]{\hat{V}_N\hat{F}_N}{\hat{V}_N}{\hat{V}_N}{\hat{F}_N}
\hat{V}_N\hat{F}_N\hat{V}_N\hat{V}_N\hat{F}_N
$ & \cincludegraphics[scale=0.4]{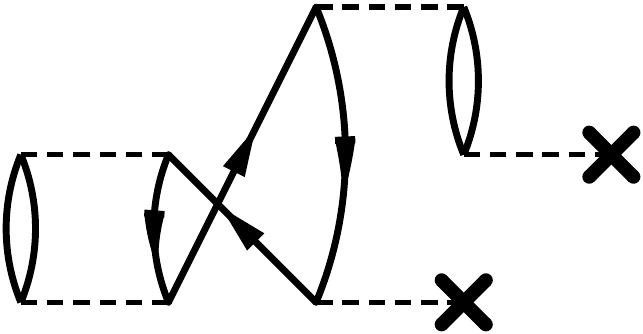} \\&&\\&&\\
E &
$
\contraction[1.0ex]{}{\hat{V}_N}{}{\hat{F}_N}
\contraction[1.0ex]{\hat{V}_N}{\hat{F}_N}{}{\hat{V}_N}
\contraction[1.5ex]{}{\hat{V}_N}{\hat{F}_N}{\hat{V}_N}
\contraction[2.0ex]{}{\hat{V}_N}{\hat{F}_N}{\hat{V}_N}
\contraction[2.5ex]{}{\hat{V}_N}{\hat{F}_N}{\hat{V}_N}
\hat{V}_N\hat{F}_N\hat{V}_N
$ & \cincludegraphics[scale=0.4]{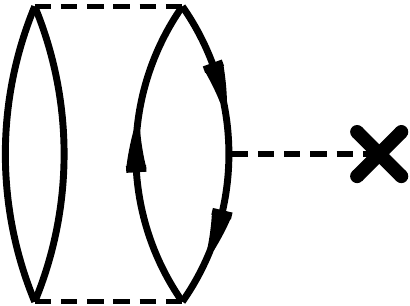} \\
F &
$
\contraction[1.0ex]{\hat{V}_N}{\hat{V}_N}{}{\hat{F}_N}
\contraction[1.5ex]{\hat{V}_N}{\hat{V}_N}{}{\hat{F}_N}
\contraction[2.0ex]{\hat{V}_N}{\hat{V}_N}{\hat{F}_N}{\hat{F}_N}
\contraction[2.5ex]{\hat{V}_N}{\hat{V}_N}{\hat{F}_N\hat{F}_N}{\hat{V}_N}
\contraction[3.0ex]{}{\hat{V}_N}{\hat{V}_N\hat{F}_N}{\hat{F}_N}
\contraction[3.5ex]{}{\hat{V}_N}{\hat{V}_N\hat{F}_N\hat{F}_N}{\hat{V}_N}
\contraction[4.0ex]{}{\hat{V}_N}{\hat{V}_N\hat{F}_N\hat{F}_N}{\hat{V}_N}
\contraction[4.5ex]{}{\hat{V}_N}{\hat{V}_N\hat{F}_N\hat{F}_N}{\hat{V}_N}
\hat{V}_N\hat{V}_N\hat{F}_N\hat{F}_N\hat{V}_N
$ & \cincludegraphics[scale=0.4]{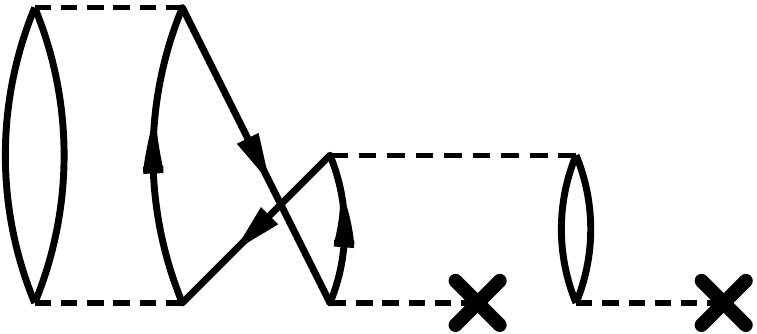} \\
\bottomrule
\end{tabular}
\end{table}

In the inevitable case of a PHL crossing (c.f. diagrams A, D, F) the latter is routed in a still easily recognizable way
such that the overall readability is not decreased. Furthermore, symmetric diagrams (if possible) are preferred over non-symmetric ones 
(c.f. diagram C). 


\section{Conclusion}
\label{conclusion}
Throughout this paper, the concept, analysis and application of the proposed automatic routing algorithm for Goldstone diagrams were illustrated. 

Section \ref{OptimizationProblem} focused on the discrete Goldstone diagram optimization problem.
A discretization of Goldstone diagrams was achieved in a grid of thirds while PHL structures were defined as a collection of PHL segments (c.f. subsections \ref{Discretization} and \ref{PHLStructure}). Suitable 
constraints for physically correct Goldstone diagrams were established (c.f. subsection \ref{Constraints}) to reduce the possible solution space. 
The remaining diagrammatic degrees of freedom (c.f. \ref{DegreesOfFreedom}) represent the parameters, that need to be optimized in order to find a well routed global minimum. 
A cost function (c.f. \ref{CostEvaluation}) evaluating the quality of individual 
Goldstone diagrams was defined.

Due to the exponentially increasing complexity of the proposed optimization problem, a genetic algorithm was developed. 
A successful crossover approach was implemented in form of a fragment-based crossover (c.f. \ref{Crossover})
transferring genetic information from two parent diagrams 
to a newly created child diagram. Different mutation processes (c.f. \ref{Mutation}) were implemented to keep 
genetic diversity throughout the algorithm and prevent premature convergence effects. 

In section \ref{Algorithmanalysis}, the convergence behavior, including a comparison of diagram complexities and the implemented 
crossover and mutation processes
were analyzed. It was shown, that Goldstone diagram solution spaces can possess enormously large complexities (c.f. \ref{ConvergenceBehaviour}) paying justice to the developed genetic algorithm 
to reach convergence within a reasonable amount of iterations.
Via the investigation of crossover and mutation processes (c.f. \ref{ReproductionandMutationRatio}), it was possible to show that the fragment-based crossover 
is beneficial by creating improved child diagrams compared to their 
individual parent diagrams. Improvements by mutation processes alone primarily occur during the final generations of the Goldstone routing algorithm. 

Section \ref{application} dealt with the applicability of the developed algorithm to different CC approaches. It was possible to reach diagrams, which fulfill all criteria 
described in subsection \ref{CostEvaluation}, in 
every conducted optimization. As a more complex testing ground, the fixpoint iterative substitution of Goldstone diagrams for the standard similarity transformed CCSD in the non-canonical post-HF procedure 
was evaluated in subsection \ref{CCSDfixpointiterations}. Additional applications of the proposed algorithm may be found in the supporting information.

\bibliographystyle{LitStyle}
\bibliography{Lit.bib}
\end{document}